\documentclass[10pt,twocolumn,letterpaper]{article}

 \usepackage[pagenumbers]{cvpr} % To force page numbers, e.g. for an arXiv version
\usepackage[dvipsnames]{xcolor}

\usepackage{epsfig}
\usepackage{graphicx}
\usepackage{amsmath}
\usepackage{amssymb}
\usepackage{algorithm}
\usepackage{algorithmic}

\usepackage{pifont}% http://ctan.org/pkg/pifont
\usepackage{color}
\usepackage{colortbl}
\usepackage[outercaption]{sidecap}
\usepackage[font=small]{caption}
\usepackage{multirow}
\usepackage{xcolor}
\usepackage{bm}
\usepackage{pythonhighlight}
\newcommand{\cmark}{\ding{51}}%
\newcommand{\xmark}{\ding{55}}%

\definecolor{Gray}{gray}{0.90}
\definecolor{white}{rgb}{1.0, 1.0, 1.0}

\definecolor{Lightgreen}{RGB}{218, 246, 230 }
\newcolumntype{a}{>{\columncolor{Lightgreen}}c}
\definecolor{Gray}{gray}{0.90}

\usepackage[algo2e]{algorithm2e}
\usepackage[dvipsnames]{xcolor}
\usepackage{graphics,color}
\usepackage{xcolor}
\definecolor{label1}{rgb}{0.76,0.59,0.77}
\definecolor{label2}{rgb}{0.28,0.5,0.72}
\definecolor{label3}{rgb}{0.33,0.63,0.36}
\definecolor{label4}{rgb}{0.79,0.4,0.17}
\definecolor{label5}{rgb}{0.94,0.53,0.2}
\definecolor{label6}{rgb}{0.72,0.86,0.59}
\definecolor{label7}{rgb}{1,1,0.65}
\definecolor{label8}{rgb}{0.93,0.62,0.61}
\definecolor{label9}{rgb}{0.4,0.15,0.33}
\definecolor{label10}{rgb}{0.75,0.21,0.29}
\definecolor{label11}{rgb}{0.35,0.73,0.8}
\definecolor{label12}{rgb}{0.94,0.9,0.32}
\definecolor{label13}{rgb}{0.96,0.76,0.48}

\newsavebox{\spleen}
\savebox{\spleen}{\textcolor{label1}{\rule{1.5in}{1.5in}}}

\newsavebox{\rkid}
\savebox{\rkid}{\textcolor{label2}{\rule{1.5in}{1.5in}}}

\newsavebox{\lkid}
\savebox{\lkid}{\textcolor{label3}{\rule{1.5in}{1.5in}}}

\newsavebox{\gall}
\savebox{\gall}{\textcolor{label4}{\rule{1.5in}{1.5in}}}

\newsavebox{\eso}
\savebox{\eso}{\textcolor{label5}{\rule{1.5in}{1.5in}}}

\newsavebox{\liver}
\savebox{\liver}{\textcolor{label6}{\rule{1.5in}{1.5in}}}

\newsavebox{\sto}
\savebox{\sto}{\textcolor{label7}{\rule{1.5in}{1.5in}}}

\newsavebox{\aorta}
\savebox{\aorta}{\textcolor{label8}{\rule{1.5in}{1.5in}}}

\newsavebox{\ivc}
\savebox{\ivc}{\textcolor{label9}{\rule{1.5in}{1.5in}}}

\newsavebox{\veins}
\savebox{\veins}{\textcolor{label10}{\rule{1.5in}{1.5in}}}

\newsavebox{\panc}
\savebox{\panc}{\textcolor{label11}{\rule{1.5in}{1.5in}}}

\newsavebox{\rad}
\savebox{\rad}{\textcolor{label12}{\rule{1.5in}{1.5in}}}

\newsavebox{\lad}
\savebox{\lad}{\textcolor{label13}{\rule{1.5in}{1.5in}}}

\newsavebox{\RV}
\savebox{\RV}{\textcolor{blue}{\rule{1.5in}{1.5in}}}
\definecolor{cvprblue}{rgb}{0.21,0.49,0.74}
\usepackage[pagebackref,breaklinks,colorlinks,citecolor=cvprblue]{hyperref}

\def\paperID{16187} % *** Enter the Paper ID here
\def\confName{CVPR}
\def\confYear{2024}

 \title{MedContext: Learning Contextual Cues for Efficient \\ Volumetric Medical Segmentation}

 \author{
  Hanan Gani$^{1}$ \quad 
  Muzammal Naseer$^{1}$ \quad 
  Fahad Khan$^{1,2}$ \quad
  Salman Khan$^{1,3}$
  \vspace{0.4em} \\
  $^{1}$Mohamed bin Zayed University of Artificial Intelligence \quad 
  $^{2}$Link\"{o}ping University \\
  $^{3}$Australian National University
}

\begin{document}
\maketitle
\begin{abstract}
Volumetric medical segmentation is a critical component of 3D medical image analysis that 
%identifies specific structural features by separating them into 
delineates different semantic regions. Deep neural networks have significantly improved volumetric medical segmentation, but they generally require large-scale annotated data to achieve better performance, which can be expensive and prohibitive to obtain. To address this limitation, existing works typically perform transfer learning or design dedicated pretraining-finetuning stages to learn representative features. However, the mismatch between the source and target domain can make it challenging to learn optimal representation for volumetric data, while the multi-stage training demands higher compute as well as careful selection of stage-specific design choices. In contrast, we propose a universal training framework called MedContext that is architecture-agnostic and can be incorporated into any existing training framework for 3D medical segmentation. Our approach effectively learns self-supervised contextual cues jointly with the supervised voxel segmentation task without requiring large-scale annotated volumetric medical data or dedicated pretraining-finetuning stages. The proposed approach induces contextual knowledge in the network by learning to reconstruct the missing organ or parts of an organ in the output segmentation space. The effectiveness of MedContext is validated across multiple 3D medical datasets and four  state-of-the-art model architectures. Our approach demonstrates consistent gains in segmentation performance across datasets and different architectures even in few-shot data scenarios.  Our code and pretrained models are available at \href{https://github.com/hananshafi/MedContext}{https://github.com/hananshafi/MedContext}.
\end{abstract}    
\section{Introduction}

\label{sec:Introduction}

Medical imaging analysis uses a variety of techniques to analyze scans obtained from procedures such as X-rays, CT scans, and MRIs \cite{LITJENS201760}. The volumetric medical segmentation of a scan is a critical component of this analysis, which identifies specific structural features by separating them into different segments or regions. A clinician may use this voxel-wise segmentation to diagnose medical conditions for treatment or further testing.  

Deep neural networks have improved volumetric medical segmentation. The convolutional encoder-decoder networks, U-NET \cite{unet2015, 3DUNET20163d}, as well as the development of vision transformers \cite{alexey2020vit}, has led to hybrid architectures \cite{karimi2021convolution, cao2021swin} with complementary strengths of self-attention and convolution for medical segmentation.  Despite the architectural advances, deep neural networks generally require large-scale annotated data to achieve better performance. However, collecting and annotating medical images at a large scale can be expensive and prohibitive due to privacy concerns.  

\begin{figure}[!t]
\centering
  \begin{minipage}{.49\columnwidth}
  	\centering
    \includegraphics[ width=\linewidth,keepaspectratio]{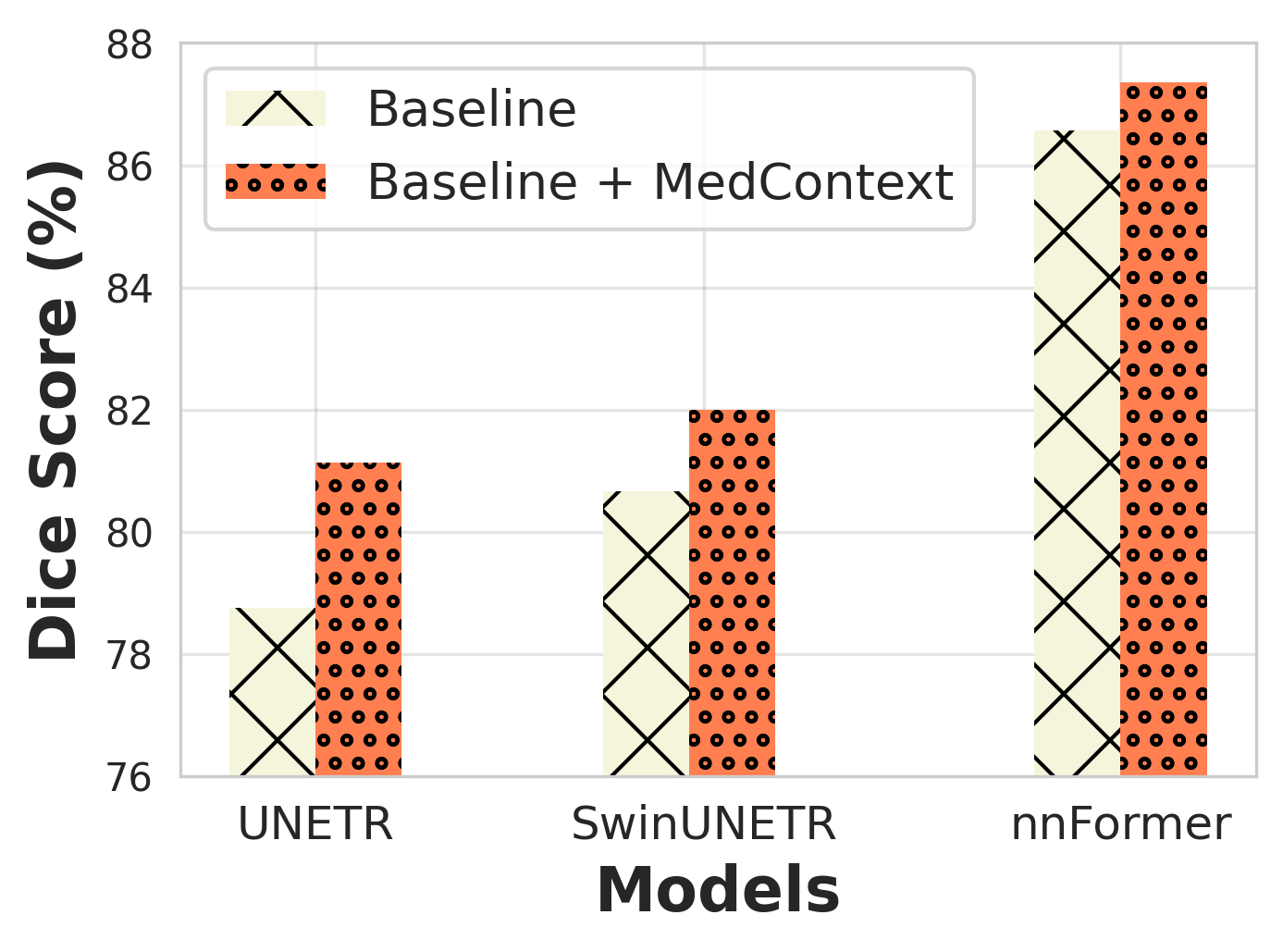}
  \end{minipage}
  \begin{minipage}{.49\columnwidth}
  	\centering
    \includegraphics[width=\linewidth, keepaspectratio]{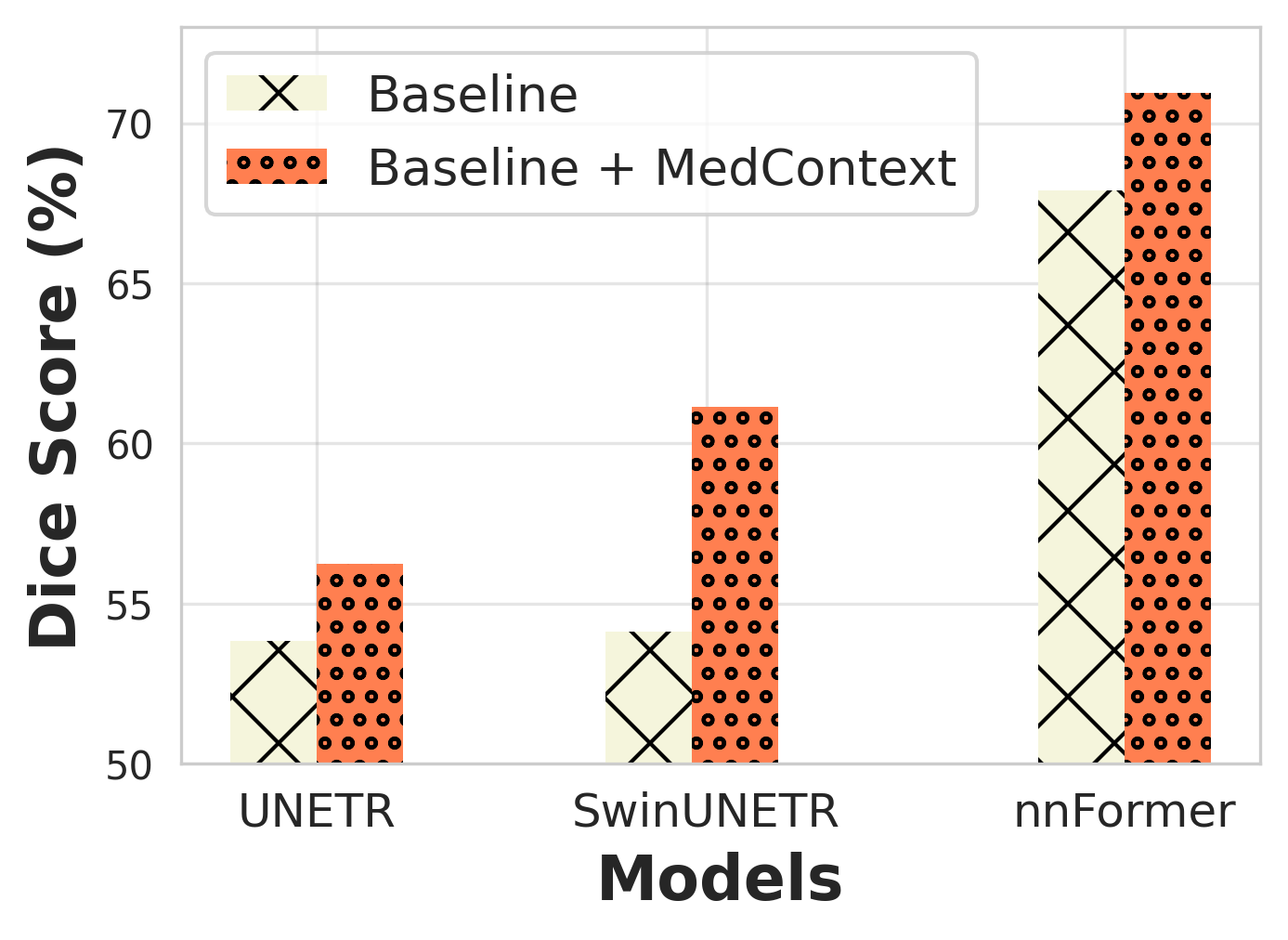}
  \end{minipage}
  \caption{Comparison, in term of Dice scores (\%), when integrating our approach into UNETR \cite{UNETR}, SwinUNETR \cite{SWIN_UNETR} and nnFormer \cite{nnFormer} for medical segmentation on Synapse dataset (Sec. \ref{sec:experiments}) using conventional setting (\textbf{Left}) and few-shot setting (5 samples only, \textbf{Right}). Without any modification to the model architecture or its training pipeline, our proposed universal approach complements the supervised voxel-wise segmentation and enhances the performance of state-of-the-art architectures. 
  %Our proposed framework demonstrates consistent gains in the few-shot dataset (5 samples only) settings (\textbf{Right}). 
  }
  \label{fig:general-and-few-shot} 
  \vspace{-1.5em}
\end{figure}

% \section{Disussion}
\begin{figure*}[!t]
\centering
  \begin{minipage}{0.48\linewidth}
  	\centering
    \includegraphics[ width=\linewidth,keepaspectratio]{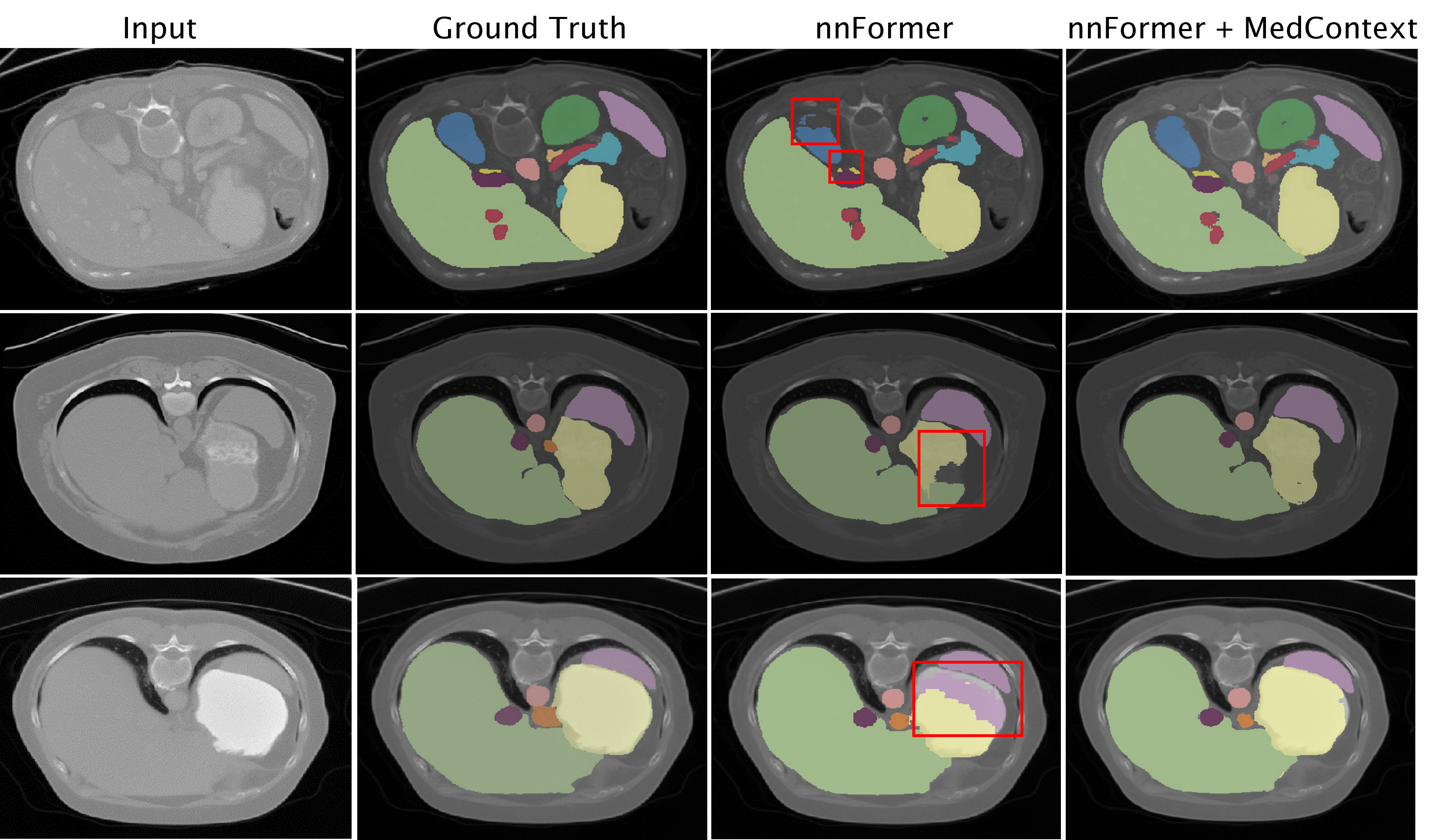}
  % \end{minipage}
  % \begin{minipage}{\linewidth}
\scalebox{0.06}{{\usebox{\spleen}}} \small Spleen ~~~\scalebox{0.06}{{\usebox{\rkid}}} \small R-Kid ~~~ \scalebox{0.06}{{\usebox{\lkid}}} \small L-Kid ~~~\scalebox{0.06}{{\usebox{\gall}}} \small Gal~~~\scalebox{0.06}{{\usebox{\eso}}} \small Eso ~~~\scalebox{0.06}{{\usebox{\liver}}} \small Liv ~~~ \scalebox{0.06}{{\usebox{\sto}}} \small Sto ~\scalebox{0.06}{{\usebox{\aorta}}} \small Aor ~~~~~~~~\scalebox{0.06}{{\usebox{\ivc}}} \small ICV ~~~~~~\scalebox{0.06}{{\usebox{\veins}}} \small PSV ~~~~~~\scalebox{0.06}{{\usebox{\panc}}} \small Pan ~ \scalebox{0.06}{{\usebox{\rad}}} \small Rad. ~\scalebox{0.06}{{\usebox{\lad}}} \small Lad.
\end{minipage}
\hfill
  \begin{minipage}{0.48\linewidth}
  	\centering
    \includegraphics[ width=\linewidth,keepaspectratio]{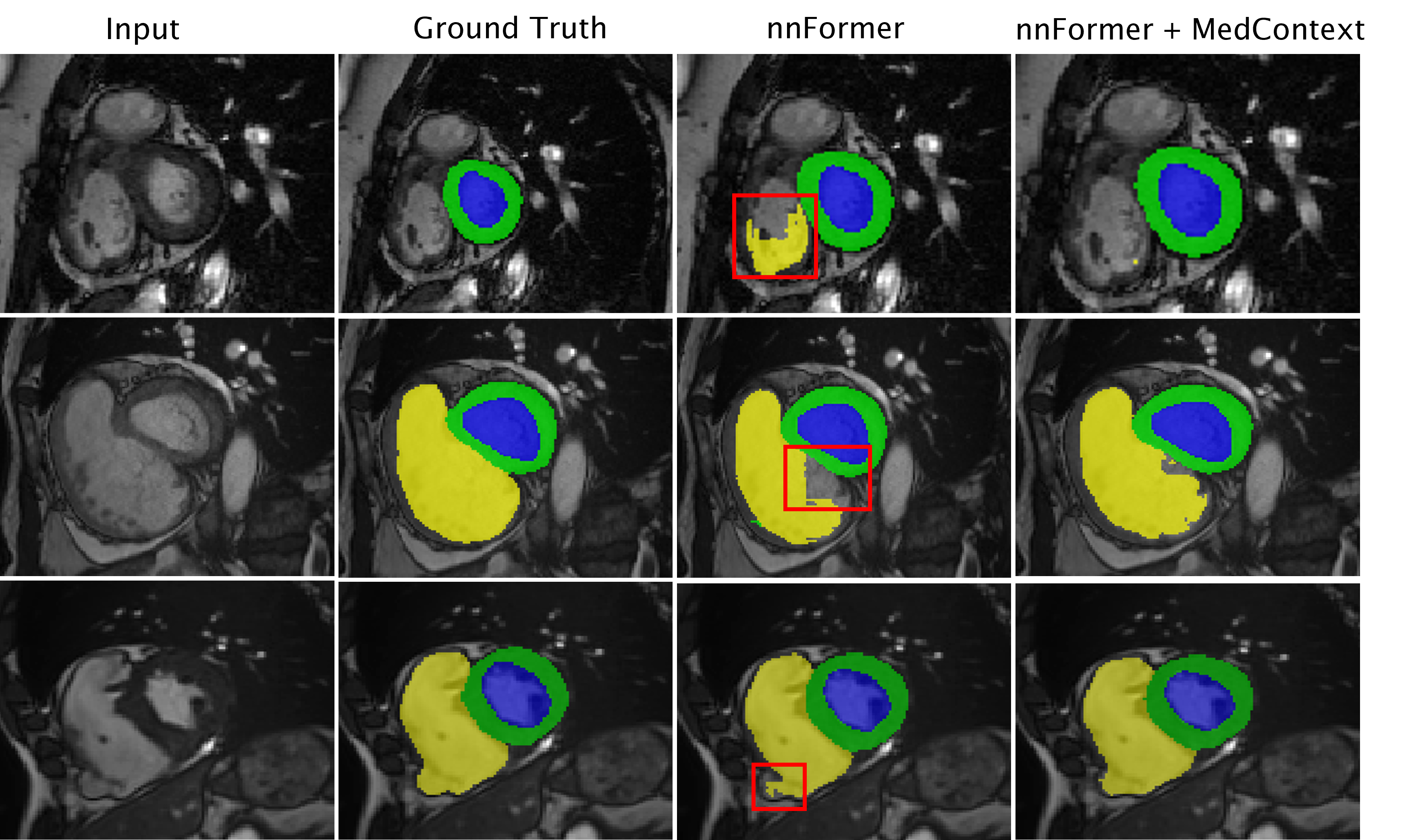}
  % \end{minipage}
  % \begin{minipage}{\linewidth}
\scalebox{0.06}{{\usebox{\rad}}} \small RV cavity ~~~~~~~~~ \scalebox{0.06}{{\usebox{\lkid}}} \small Myocaridum ~~~~~~~~~~~~
\scalebox{0.06}{{\usebox{\RV}}} \small LV cavity 
\end{minipage}
  \caption{Qualitative Comparison between the baseline nnFormer \cite{nnFormer} and our proposed MedContext integrated with nnFormer. 
  % We showcase the benefit of our 3D-MSR framework implemented on the nnFormer architecture. 
  The examples display different abdominal organs (Synapse) (\textbf{Left}) and regions of the heart (ACDC) (\textbf{Right}), with their corresponding labels in the legend below. The baseline nnFormer struggles to accurately segment the organs and heart regions. In certain cases, it gives false segmentation results highlighted in red boxes. Best viewed zoomed in. Refer to supplementary material for additional qualitative comparisons. }
  % \caption{Qualitative comparison on multi-organ synapse dataset: We  showcase the benefit of our 3D-MSR framework implemented on the nnFormer architecture. The images display various abdominal organs, with their corresponding labels in the legend below. The existing baseline method struggles to accurately segment the organs as can be seen from the red boxes. Best viewed in zoom. Further results can be found in the supplementary material. }
  \label{fig:qualitative_synapse_acdc_nnformer}
\end{figure*}

To deal with the data scarcity, weights learned on ImageNet \cite{deng2009imagenet} can be used to initialize the encoder, however, pre-training on 2D natural images may not capture the contextual information essential to understanding 3D medical images. Recent studies \cite{SWIN_UNETR, hatamizadeh2022unetformer, xie2021unified} explore self-supervised pre-training on extra auxiliary medical data, but this approach has two limitations: a) the training process consists of a two-stage pre-training process on the auxiliary data, followed by fine-tuning on the target data {which can be computationally expensive},
and b) the success of fine-tuning depends on how well the auxiliary data distribution matches the target data. Moreover, there may not be a direct relationship between the self-supervised objectives (e.g., inpainting, solving jigsaws, or predicting rotation) and voxel-wise segmentation. Therefore jointly optimizing such self-supervised losses with 3D segmentation is non-trivial.

To address these limitations, we propose a generic training framework dubbed \textit{MedContext} to learn self-supervised contextual cues jointly with supervised voxel segmentation without requiring  large-scale annotated volumetric medical data. Specifically, we propose to reconstruct the masked organs or parts of an organ of an input image in the output segmentation space. Since our voxel-wise segmentation reconstruction is well aligned with the voxel-wise prediction task, both tasks can be optimized together. To further reduce the disparity between voxel segmentation reconstruction and prediction tasks, we deploy a student-teacher distillation strategy to guide reconstruction from a slow-moving online teacher model which also helps avoid representation collapse. MedContext encourages contextual learning within the model and allows it to learn local-global relationships between different input components. This leads to better segmentation of organ boundaries (see Fig. \ref{fig:qualitative_synapse_acdc_nnformer}).

Our proposed approach is architecture-agnostic  and can be incorporated into any training framework, making it universally applicable. We integrate our approach into three recent state-of-the-art medical 3D transformer based architectures: UNETR \cite{UNETR}, SwinUNETR \cite{SWIN_UNETR} and nnFormer \cite{nnFormer}; and one CNN based 3D architecture PCRLv2 \cite{zhou2023pcrlv2}. Using these architectures, we validate our approach across three medical imaging datasets: Multi-organ Synapse \cite{BTCV}, ACDC \cite{ACDC} and BraTS \cite{brats17, baid2021rsna}. We observe consistent performance gains on all the datasets.
% with our approach. 
Specifically, on the Synapse dataset with nnFormer and UNETR architectures (Fig. \ref{fig:general-and-few-shot}), we observe a performance improvement of $\sim$1\% and 2\% respectively in terms of Dice score. Similarly, on ACDC dataset with UNETR architecture, we observe a $\sim$4\% gain in the Dice score compared to the baseline. We further assess the effectiveness of our framework in comparison to various pretraining-finetuning methods including \cite{chen2020simple}, \cite{xie2022simmim}, and \cite{he2022masked}. Our evaluation reveals consistent performance improvements across all compared methods.

In summary, our contributions are three-fold: %\vspace{-0.5em}
\begin{enumerate}\setlength{\itemsep}{0em}

 \item We propose a universal training framework to jointly optimize supervised segmentation 
 % prediction 
 and self-supervised segmentation reconstruction via student-teacher knowledge distillation for the volumetric medical data. 

\item Our approach induces contextual knowledge in the network by learning to reconstruct the missing organ or parts of an organ in the output segmentation space.

\item  We validate the effectiveness of our approach across multiple 3D medical datasets and state-of-the-art model architectures. Our approach complements existing methods and improves segmentation performance in conventional as well as few-shot data scenarios.
\end{enumerate}

\section{Related Work}
% In this section, we discuss the existing works related to 3D medical segmentation, Knowledge distillation and Mask Image modeling. 
\textbf{3D Medical Segmentation:} Several U-Net \cite{unet2015} based encoder-decoder architectures have been proposed to solve the problem of 3D medical segmentation. {\cite{cciccek20163d}} modifies the basic U-Net architecture by replacing 2D operations with 3D operations to capture the 3D context. \cite{nnUNet} proposed to learn multi-scale feature representations from varying resolutions for multi-organ segmentation. Other works have suggested integrating the contextual information with CNN-based frameworks using image pyramids ~\cite{zhao2017pyramid}, large kernels ~\cite{peng2017large}, dilated convolution ~\cite{chen2018encoder}, and deformable convolution ~\cite{li2020pgd}. 
% Transformer \cite{alexey2020vit} based approaches have recently gained popularity due to their tendency to encode global information owing to the presence of self-attention modules. 
Few recent works \cite{karimi2021convolution, cao2021swin} have explored the use of transformer architectures for 3D volumetric segmentation by dividing the volumetric images into 3D patches which are then flattened to construct a 1D embedding and passed to transformer module. More recently, hybrid architectures ~\cite{TransFuse, valanarasu2021medical, TransUNet, lin2022ds, UNETR, nnFormer} combining the strengths of both CNNs and transformers have been proposed to encode both local and global contexts. We build our approach on these hybrid architectures by proposing a complementary training mechanism that induces contextual information in these architectures. 

\textbf{Masked Image Modeling (MIM): } MIM has emerged as an appealing self-supervised representation learning method, fueled by the success of ViTs. Recent MIM methods \cite{he2022masked, xie2022simmim} are trained to predict the pixel values to reconstruct the corresponding missing tokens in the input. Recent works on videos \cite{tong2022videomae,he2022masked, feichtenhofer2022masked} use temporal masking to learn self-supervised representations for videos. In case of medical imaging, \cite{Chen_2023_WACV} explores MIM on 3D medical data. \cite{zhou2023pcrlv2} proposes pixel restoration task on 3D medical images. However, these approaches are based on a two-stage process of self-supervised pretraining on large public datasets \cite{Harmon2020ArtificialIF, SETIO20171} and then finetuning. Different from existing self-supervised learning methods which
require an additional stage of supervised finetuning on labeled data, we synergically incorporate
MIM with the voxel-wise segmentation objective in a single training framework without any external data.

\begin{figure*}[t]
  \centering
    \includegraphics[width=0.96\linewidth]{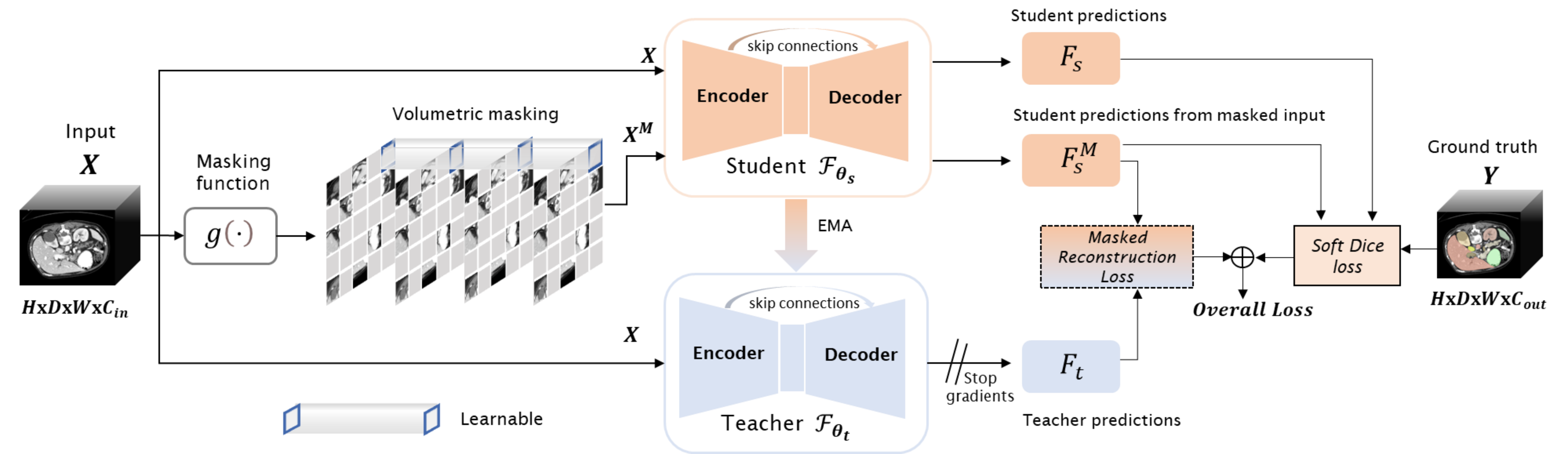}
    %\vspace{-0.45em} 
    \caption{\emph{Overview of our MedContext approach:} The original 3D volume is masked and fed to the student model (top-row) along with the original input. The teacher model (bottom-row) is only fed with the original volume. The difference between the semantic voxelwise  predictions for the masked and original inputs corresponding to the student and teacher networks respectively is minimized to guide the reconstruction of masked regions in the output segmentation space. Our approach induces contextual consistency by enabling the model to reconstruct and segment the missing organs or organ parts and therefore yields more precise and accurate segmentation results. 
    % We show this framework is universal in nature and can be applied to any existing 3D  segmentation approach to boost results.
    }
    
    %in the model and builds relationships between objects and enables the network to capture intricate relationships between various objects in the input volume, therefore yielding more precise and accurate segmentation results.   }
    \label{fig:main_diagram}

\end{figure*}

\textbf{Knowledge Distillation: }Introduced by \cite{hinton2015distilling}, knowledge distillation aims to transfer knowledge from a complex model (teacher) to a simpler model (student). Recent works \cite{richemond2020byol, caron2021emerging} have explored the use of knowledge distillation with an online student-teacher strategy for self-supervised learning. \cite{richemond2020byol} employs a student-teacher based knowledge distillation framework for self-supervised representation learning using a contrastive objective. \cite{caron2021emerging} uses a dynamic student-teacher framework to learn local-global correspondences by distilling the knowledge from the teacher which is built dynamically. \cite{qin2021efficient} extends the knowledge distillation approach to medical segmentation where a strong teacher model is used to distill knowledge within a small compressed model. All these approaches are applicable in the 2D data scenario. In contrast, our proposed knowledge distillation framework works on 3D volumetric inputs and serves as a means to reconstruct the missing voxels in the segmentation space in a single end-to-end training stage. 
\section{Methodology }
The existing 3D medical segmentation methods focus on optimizing voxel-wise segmentation, typically by minimizing the dice loss between predicted and ground truth segmentation masks. 
However, this approach may not capture the underlying 3D structure of the input or the contextual relationships between different objects, especially in data-scarce scenarios such as in the medical domain. To address this limitation, we propose a novel approach that learns contextual relationships between different organs or organ parts in the output segmentation space. Specifically, we formulate the problem as reconstructing missing organ parts from a masked input volume in the segmentation space using student-teacher knowledge distillation.

Our proposed approach complements the voxel-wise segmentation task by inducing contextual consistency through joint optimization of two objectives: \textbf{a)} voxel-wise segmentation reconstruction from the masked input, and \textbf{b)} supervised voxel-wise segmentation prediction. 
The learning objectives enable the model to reconstruct and segment the missing organs or organ parts, thereby encoding contextual information. This allows the model to learn \textit{local-global relationships} between different input components for a data-efficient and accurate 3D semantic segmentation.
In summary, our approach provides a solution to the limitations of current 3D medical semantic segmentation methods by incorporating contextual learning via joint optimization of multi-task objectives.
Next, we explain the network architecture, objective functions and optimization scheme.

\subsection{Architecture}
\label{sec: Architecture}
Our approach is complementary and can be applied to the existing encoder-decoder architectures designed for 3D medical image segmentation. Specifically, we show the benefit of our approach using three recently introduced models: UNETR \cite{UNETR}, SwinUNETR \cite{SWIN_UNETR}, and nnFormer \cite{nnFormer}.  %We use 3D U-Net \cite{cciccek20163d} based encoder-decoder architectures in our work. 
In these architectures, the encoder processes a 3D input volume and produces latent feature representation. The decoder then maps this representation to the corresponding segmentation map through upsampling blocks. Additionally, skip connections are used to exchange the multi-level features across different encoder-decoder layers. 
We illustrate how our approach can be integrated into these hybrid architectures, which combine both convolution and transformer components, for 3D medical image segmentation.

As shown in Fig.~\ref{fig:main_diagram} our design includes a student $\mathcal{F}_s$ and a teacher $\mathcal{F}_t$ network that operate on the input volume $\bm{X} \in \mathbb{R}^{H\times W \times D}$ and its masked version $\bm{X}^M \in \mathbb{R}^{H\times W \times D}$ generated using the masking function $g(.)$. Here, $H$, $W$, and $D$ represent the height, width, and depth of the 3D input volume, respectively. 
During the training phase, the input views are fed to the student-teacher framework as 3D patches, which generates voxel-wise semantic logits for each input view. The student network is provided with both the masked ($\bm{X}^M$) and unmasked ($\bm{X}$) inputs, and the corresponding output voxel-wise semantic logits are denoted as $\bm{F}_{s}$ and $\bm{F}_{s}^{M}$, respectively. On the other hand, the teacher network is provided with the original unmasked input $\bm{X}$ which outputs voxel-wise semantic logits denoted as $\bm{F}_{t}$. The feature map produced at intermediate layers of the 3D architecture has a shape of $\frac{H}{P_1} \times \frac{W}{P_2} \times \frac{D}{P_3} \times C $, where $(P_1, P_2, P_3)$ is the resolution of each patch and $C$ is the feature dimension.

For each output prediction from the student , a supervised loss is computed using the ground truth label $\bm{Y}$, as shown in the figure. Additionally, a self-supervised objective is minimized between the masked student logits $\bm{F}_{s}^{M}$ and the teacher logits $\bm{F}_{t}$. Finally, both the supervised and self-supervised objectives are jointly optimized during our single-stage training process.

\subsection{Volumetric Masking Strategy}
\label{subsec: Volumetric Masking Strategy}
To model contextual relationships, we employ a masking technique on the original input $\bm{X}$ to reconstruct missing parts in the segmentation space. As illustrated in Fig. \ref{fig:main_diagram}, we ensure mask consistency across the depth by applying the same mask to all subsequent slices in the volume, thereby ensuring that any masked organ or region in the first slice remains masked in all subsequent slices. Our encoder-decoder architectures convert the input volume into patch tokens before processing. To generate a masked view $\bm{X}^M$, we randomly mask a certain fraction $\delta$ of the patch tokens. Following \cite{devlin2018bert}, the masked tokens are replaced with learnable tokens $\mathcal{H}_\xi$, such that,
\begin{equation}
\label{eqn:masking-strategy}
\bm{X}^M = g(\bm{X}, \delta) = \bm{X} \circ (1 - \bm{I}_\delta) + \mathcal{H}_\xi \circ \bm{I}_\delta ,
\end{equation}
where $\bm{I}_\delta$ is a binary mask generated according to a Bernoulli distribution using $g(.)$, i.e., $\bm{I}_\delta \sim$ Bernoulli($\delta$) and $\circ$ denotes the element-wise product. This masking technique prevents information leakage from neighboring cubes by enforcing the same masking map for all cubes along the depth dimension. Our approach encourages the model to learn contextual semantic relationships by recovering the segmentation map for such masked inputs.

\subsection{Voxel-wise Segmentation Reconstruction} %from the Masked Input
\label{subsec:Voxelwise Segmentation Reconstruction from the Masked Input}

Using masked input, we reconstruct the segmentation map to facilitate learning contextual semantic relationships. To accomplish this task, a naive approach would be to feed both the original and masked views into a single model and optimize the outputs of both views using supervised loss with ground truth labels. However, this approach has certain limitations. When a single model is used to reconstruct the masked volume, the model's weight updates are solely dependent on information from the current views. A number of self-supervised learning techniques, such as DINO \cite{caron2021emerging} and BYOL \cite{richemond2020byol}, have shown that leveraging the knowledge acquired by the model during previous weights updates can lead to more effective guidance and avoids representation collapse. This can be achieved by implementing a student-teacher strategy where teacher weights are updated by a moving average of the student weights. In this way, the collective knowledge learned during previous weight updates assists in the reconstruction of the masked views and thereby induces enriched contextual cues. Additionally, the teacher network provides soft semantic targets that guide the training of the student network. The soft targets contain information about the relationships between the views and aids the student network in generalizing more effectively. 

We initialize the student model $\mathcal{F}_s$ and teacher model $\mathcal{F}_t$ with the same randomly initialized weight parameters. The student network receives both the original and masked inputs, while the teacher network receives only the original non-masked input. The networks generate voxel-wise semantic logits, represented by \{$\bm{F}_s^{M}$, $\bm{F}_s$\} and $\bm{F}_t$ respectively, which are processed to obtain semantic voxel-wise segmentation map predictions. Following this, we reconstruct semantic voxel-wise logits of the masked input  from the student model guided by two supervised signals: \emph{supervision through knowledge distillation} and \emph{ground truth labels}, as explained below. \\

\noindent
\textbf{Reconstruction through Knowledge Distillation:} A self-supervised distillation loss (Eq. \ref{eqn:distillation-loss}) is used to guide the training of the student network to encourage modeling the contextual consistency. It minimizes the difference between the voxel-wise logits generated by the teacher network  given the original input $\bm{F}_{t}$ and the voxel-wise logits produced by the student network using the masked input $\bm{F}_{s}^M$. The objective function, referred to as Consistency Loss (CL), is denoted as $\mathcal{L}_{c}(\bm{F}_{s}^{M},\bm{F}_{t})$ and is expressed as,
\begin{equation}
\label{eqn:distillation-loss}
    \mathcal{L}_{c}(\bm{F}_{s}^{M},\bm{F}_{t}) = \frac{\Vert\ \bm{F}_{s}^{M} - \bm{F}_{t}  \Vert\ _{2}^{2}}{\Vert\ \bm{F}_{t}  \Vert\ _{2}^{2}}.
\end{equation}

\noindent
\textbf{Reconstruction through Ground truth Labels:} The voxel-wise semantic logits output by the student network for a given masked input $\bm{F}_{s}^M$ are further reconstructed using the ground truth labels. This is accomplished by minimizing the soft dice loss \cite{milletari2016v} using the ground truth labels $\bm{Y}$. The general expression for Dice-CE Loss for some arbitrary output prediction $\bm{F}$  is given as,
\begin{align}
\label{eqn:dice-ce-loss}
    \mathcal{L}_{Dice-CE}(\bm{Y},\bm{F}) =  1-\sum_{c=1}^{C}\ & \left(\frac{2*\sum_{v=1}^{V} \bm{Y}_{v,c} \cdot {\bm{F}_{v,c}}}{\sum_{v=1}^{V}\bm{Y}_{v,c}^2 + \sum_{v=1}^{V}{\bm{F}_{v,c}^2}} \right. \notag\\ 
    & +  \left. \sum_{v=1}^{V}\bm{Y}_{v,c} \log {\bm{F}_{v,c}} \right),
\end{align}
where, $C$ denotes the number of classes;  $V$ denotes the number of voxels; $\bm{Y}_{v,i}$ and ${\bm{F}_{v,i}}$ denote the ground truths and output probabilities for class $i$ at voxel $v$, respectively.
In our case, the supervised objective function for reconstruction is calculated using above Dice-CE loss between the ground truth label $\bm{Y}$ and voxel-wise semantic logits output $\bm{F}_{s}^M$ and is denoted as  $\mathcal{L}_{Dice-CE}(\bm{Y},\bm{F}_s^M)$ and referred to as Masked Student Loss (MSL).

Our reconstruction objective on the masked input volumes induces contextual consistency to enhance segmentation performance. The reconstruction of the missing regions enables the network to capture intricate relationships between various organs in the input volume and learn the broader context of the input volume going beyond the local features to segment organs. 
This %strategy 
promotes the preservation of the global structure of the input volume, yielding more precise and accurate segmentation results.
Moreover, the objective of predicting missing regions in the masked input further encourages the model to learn correspondence between the local and global structure in the input volume. Masked input volumes represent local views of inputs that are matched with original global input volumes. This matching objective allows the model to capture the relationships between neighboring regions which is essential for learning class-specific semantic features to better capture the object boundaries and shapes.

\subsection{Supervised Voxel-wise Segmentation}
% Prediction}
\label{supervised-segmentation}
Our primary task of supervised voxel-wise segmentation prediction takes place in conjunction with the voxel-wise segmentation reconstruction as discussed above. For the supervised voxel-wise segmentation, we optimize the predictions of the original input $\bm{X}$ from the student network $\bm{F}_s$ through the supervision of the ground truth labels $\bm{Y}$ using Soft Dice Loss (Eq. \ref{eqn:dice-ce-loss}) denoted by the objective $\mathcal{L}_{Dice-CE}(\bm{Y},\bm{F}_s)$.
%, which acts as a regularizer to prevent overfitting and mitigate the risk of mode coll apse. 
% We provide an ablative analysis of our different losses  in Sec. \ref{subsec:Ablation Studies}.

\subsection{Overall Multi-task Objective}
\label{loss-func}
Our framework leverages a combination of supervised and self-supervised losses to optimize the learning process. The multi-task objectives synergistically reinforce each other and provide complementary advantages. The overall loss objective $\mathcal{L}$ is given as,
\begin{align}
\label{eqn:combined-loss}
         \mathcal{L} =   & \mathcal{L}_{Dice-CE}(\bm{Y},\bm{F}_s)  +  \mathcal{L}_{Dice-CE}(\bm{Y},\bm{F}_s^M) \nonumber \\
     & + \beta \mathcal{L}_{c}(\bm{F}_s^M, \bm{F}_t) ,
\end{align}
where the hyperparameter $\beta$ controls the contribution of self-supervised consistency loss during optimization. 

\subsection{Optimization strategy}
\label{optimization}
Following a typical student-teacher optimization strategy as ultilized by \cite{grill2020bootstrap, caron2021emerging}, the gradient of the total loss is backpropagated through the student network  and parameters are updated as follows,
\begin{align}
\label{eqn:grad-descent}
         \Theta \leftarrow \Theta - \alpha \cdot \nabla_\Theta (\mathcal{L}),
\end{align}
where $\Theta$ represents the joint parameters of student network ($\theta_{s}$) and learnable mask embeddings ($\mathcal{\xi}$) i.e. $\Theta = \{\theta_{s}; \mathcal{\xi}\}$. The teacher network is updated via exponential moving average (EMA) of the weights of the student network using,
\begin{align}
\label{eqn:ema}
         \theta_{t} \leftarrow \lambda \theta_{t} +  (1 - \lambda)\theta_{s},
\end{align}
where $\theta_{t}$ denote the parameters of teacher and, $\lambda$ follows the cosine schedule from 0.996 to 1 during training. The gradient step through the student network comprises of the contributions from both the supervised and self-supervised objectives, thereby aiding  in the reconstruction of the masked input by updating the differentiable volumetric embeddings associated with the masked regions in the input volume.\\ 
\textbf{Avoiding Mode collapse:} The student-teacher frameworks in general are often prone to mode collapse where the student model fails to learn the entire range of outputs that the teacher model can generate. Our framework avoids the mode collapse by introducing additional supervised loss from the student corresponding to the output from the masked volume, which in addition to assisting in the masked reconstruction, also acts as a discriminative objective to encourage the student model to learn a more diverse set of outputs that better match the full range of outputs generated by the teacher. Additionally, it further encourages teacher model to generate high-quality targets for distillation. We show the effect of our different losses in Sec. \ref{subsec:Ablation Studies}.
\section{Experiments}
\label{sec:experiments}

\begin{table*}[!ht]
      \centering \small
      % \caption{\small Comparison with state-of-the-art baselines on abdominal multi-organ Synapse dataset.: Our 3D-MSR achieves better segmentation performance compared to the existing state-of-the-art baselines. The performance is measured in terms of Dice score and HD95. Best results are shown in bold.}
      % \vspace{-0.5em}
      \setlength{\tabcolsep}{6.5pt}
      \scalebox{0.85}[0.85]{
        \begin{tabular}{l c | c c  c c c c c c|aa}
        \toprule
            \multirow{2}{*}{Models} & \multirow{2}{*}{\rotatebox[origin=c]{0}{MedContext}}  & \multirow{2}{*}{\rotatebox[origin=c]{0}{Spleen}} &  \multirow{2}{*}{\rotatebox[origin=c]{0}{Right Kidney}} &  \multirow{2}{*}{\rotatebox[origin=c]{0}{Left Kidney}} & \multirow{2}{*}{\rotatebox[origin=c]{0}{Gallbladder}}  & \multirow{2}{*}{\rotatebox[origin=c]{0}{Liver}}  & \multirow{2}{*}{\rotatebox[origin=c]{0}{Stomach}} & \multirow{2}{*}{\rotatebox[origin=c]{0}{Aorta}} &  \multirow{2}{*}{\rotatebox[origin=c]{0}{Pancreas}} &  \multicolumn{2}{c}{Average} 
            
            \\ \cmidrule{11-12}
             & & & & & & & & & & HD95 $\downarrow$ & DSC $\uparrow$ \\
                \midrule
                \midrule

        \multirow{2}{*}{UNETR} & \xmark &89.64 & 83.02 & 84.86 & 63.06 & 95.58 & 73.06 & 87.47 & 53.40 &  11.04 & 78.76  \\
          & \cmark &90.73  &83.36  & 86.03 & 67.94  & 95.59  & 78.62  &87.30  & 59.51  & \textbf{9.44}  &\textbf{ 81.13} \\
        \midrule
        
        \multirow{2}{*}{Swin-UNETR}  & \xmark&86.33  &80.63  &84.07 &67.24  &94.98  &74.97  &90.53 &66.49  &20.32   &  80.66
        \\

        % & \cmark & 87.22 &81.33  &83.55  &69.06 &95.31  &73.90  &90.10   &73.98  & \textbf{37.72} & \textbf{81.80} \\
        & \cmark & 91.45 &80.80  &84.85  &67.70 &94.60  &76.20  &90.88   &67.74  & \textbf{14.45} & \textbf{82.00} \\
        \midrule
        \multirow{2}{*}{nnFormer}  & \xmark & 90.51 & 86.25 & 86.57 & 70.17 &96.84 & 86.83  & 92.04 &  83.35&  10.63 & 86.57\\
        & \cmark & 95.97 &87.05  &87.63 &72.87  &96.43 &84.57   &91.85 & 82.40   &\textbf{8.29}   &\textbf{ 87.35}\\
        
        \bottomrule
        \end{tabular}
        }
        \caption{\small Abdominal multi-organ Synapse: Our MedContext consistently improves the segmentation performance of all organs across different models. We observe significant improvements in HD95 along with the dice score (DSC). The best results  are highlighted in bold.}
% \vspace{-0.2cm}
    \label{table:synapse}
\end{table*}

%%%% ACDC Results
\begin{table*}[t]
\begin{minipage}{0.34\textwidth}
 \centering \small
 % \caption{\small Dice score (\%) for ACDC Comparision: We report the performance on the right ventricle (RV), left ventricle (LV), and myocardium (MYO) along with mean results using Dice metrics. Our 3D-MSR outperforms the baselines in all the cases. Best results are shown in bold.}
  \vspace{-0.5em}
        \setlength{\tabcolsep}{3pt}
        \scalebox{0.75}[0.75]{
        \begin{tabular}{lc|ccc|a}
        \toprule
       \rowcolor{white}
        {Models} & {MedContext} & RV & Myo & LV & Average  \\
                \midrule
                \midrule
        \multirow{2}{*}{UNETR} &\xmark &77.81  &72.74  &79.46  & 76.67 \\
       &\cmark &  84.77 &75.82  & 81.21  & \textbf{80.60}  \\
        \midrule
        
         \multirow{2}{*}{SwinUNETR} &\xmark & 83.47 & 75.54 & 83.09 & 80.70 \\
        &\cmark  &84.79 &79.17 &86.15 &\textbf{83.38}    \\
        \midrule
        % Swin-UNETR & xx.xx & xx.xx & xx.xx & xx.xx \\
        \multirow{2}{*}{nnFormer} &\xmark & 91.18 & 86.24 & 94.07 & 90.50 \\
        % nnFormer 92.06 90.94 89.58 95.65
        &\cmark &92.14 & 86.52  &93.52  & \textbf{90.73} \\
        \bottomrule
        \end{tabular}}
    \caption{\small ACDC: We report the performance on the right ventricle (RV), left ventricle (LV), and myocardium (MYO).
  % Our 3D-MSR outperforms the baselines in all cases.
  }
    \label{table:ACDC}
\end{minipage}
\hfill
\begin{minipage}{0.34\textwidth}
    \centering \small
    \vspace{-0.5em}

        \label{table:BraTS}
        \setlength{\tabcolsep}{3pt}
        \scalebox{0.75}[0.75]{
        \begin{tabular}{lc|ccc|a}
        \toprule
        \rowcolor{white}
        Models & MedContext & WT & ET & TC & Average  \\
                \midrule
                \midrule
        \multirow{2}{*}{UNETR} &\xmark &87.35  &90.88  &84.29  & 87.50 \\
        &\cmark &87.43 &91.45  &85.23  & \textbf{88.04}  \\
        \midrule
        
        \multirow{2}{*}{SwinUNETR} &\xmark &90.36  &91.72  &86.24 & 89.44 \\
        &\cmark &90.57 &92.30 &86.64 & \textbf{89.83}    \\
        \midrule
        % Swin-UNETR & xx.xx & xx.xx & xx.xx & xx.xx \\
         \multirow{2}{*}{nnFormer} &\xmark &80.80  &58.86  & 77.42 &72.36  \\
        &\cmark & 81.00  &59.87  &77.45 & \textbf{72.78} \\
        \bottomrule
        \end{tabular}}
    \caption{\small BraTS: We report the performance on three brain tumour types, demonstrating effectiveness of our approach for all the three cases.
    % the effectiveness of our approach on a relatively larger BraTS dataset. 
    % Our 3D-MSR improves the baseline models for all cases.
    }
    \label{table:brats}
\end{minipage}
\hfill
\begin{minipage}{0.28\textwidth}
\small \centering
\vspace{-0.5em}
    % \caption{\small Few-shot learning performance: We demonstrate the label-efficiency property of our 3D-MSR by reporting  5-sot performance on multi-organ synapse \cite{BTCV} and ACDC \cite{ACDC} datasets across all the architectures. Best results are shown in bold.}
    % We report the 5-sot performance on multi-organ synapse \cite{BTCV} and ACDC \cite{ACDC} datasets. Our approach when complemented the baselines shows a higher performance in terms of Dice metric. Best results are shown in bold.

     \setlength{\tabcolsep}{3pt}
        \scalebox{0.75}[0.75]{
        \begin{tabular}{lc|aa}
        \toprule
        \rowcolor{white}
        Models & MedContext&Synapse &ACDC\\
        \midrule
        \midrule

        \multirow{2}{*}{UNETR} &\xmark    &53.83 & 18.53 \\
        &\cmark & \textbf{56.25} & \textbf{28.63}\\
        \midrule
        \multirow{2}{*}{SwinUNETR} &\xmark &54.13  &32.62 \\
        &\cmark & \textbf{61.15} & \textbf{35.80}   \\
        \midrule

        \multirow{2}{*}{nnFormer} &\xmark&67.90 &52.23 \\
       &\cmark & \textbf{70.96}  & \textbf{58.05 }  \\
        \bottomrule
        \end{tabular}}

    \caption{\small Few-shot: Performance of our MedContext in 5-shot scenario (5 samples only).
    }
    \label{table:few-shot}
\end{minipage}%\vspace{-1em}
\end{table*}

\textbf{Datasets:}
% \label{subsec:datasets}
We evaluate on three volumetric medical scan datasets. %Specifically we use the respective training pipeline of each of the architecture to validate the effectiveness of our approach and justifying the claim that our approach is agnostic to architecture and training pipeline. We provide the details of each of the datasets as follows:\\
% \noindent
%
% \noindent \textbf{Synapse BTCV Multi-organ Dataset \cite{BTCV}:}  
\textbf{Synapse BTCV Dataset}:  
The BTCV dataset \cite{BTCV}, known as Synapse for Multi-organ CT Segmentation,
%is derived from the MICCAI Multi-Atlas Labeling Beyond the Cranial Vault challenge. It
includes abdominal CT scans of 30 subjects encompassing 8 organs.
% The dataset is expertly annotated under the supervision of clinical radiologists at Vanderbilt University Medical Center. Each scan is captured using contrast enhancement in the portal venous phase and contains between 80 to 225 slices with 512×512 pixels. The thickness of each slice varies from 1 to 6 mm. 
Following previous methods, we adopt the same dataset split as used in \cite{TransUNet} with 18 train samples and test on the remaining 12 cases. We evaluate the performance on eight abdominal organs (i.e. spleen, right kidney, left kidney, gallbladder, liver, stomach, aorta, and pancreas) using Dice Similarity Coefficient (DSC) and 95\% Hausdorff Distance (HD95).
In all the cases, the intensities of input volumes are normalized from the range of [-1000, 1000] to [0,1] Hounsfield Units (HU). 
% For \textit{UNETR}  \cite{UNETR} and \emph{SwinUNETR} \cite{SWIN_UNETR}, we follow their respective data processing pipeline where each volume is pre-processed independently and resampled to have an isotropic voxel spacing of [1.5, 1.5, 2.0]. Input is sampled at a crop size of 96 x 96 x 96. For \textit{nnFormer} \cite{nnFormer}, we use their data preprocessing pipeline in which each CT scan is independently processed by applying patch cropping to sample the input at a resolution 128 × 128 × 64 with a spacing of [0.76, 0.76, 3].
%
\textbf{ACDC Dataset:} The ACDC dataset \cite{ACDC} is a collection of cardiac MRI images and associated segmentation annotations for the right ventricle (RV), left ventricle (LV), and myocardium (MYO) of 100 patients, obtained from actual clinical exams. we split the dataset into 80 training and 20 testing samples following \cite{nnFormer} and report the results on all three classes using Dice similarity coefficient (DSC).
% The dataset includes patients with various heart conditions, including normal patients, patients with a myocardial infarction, dilated cardiomyopathy, hypertrophic cardiomyopathy, and abnormal right ventricle. 
% For  \textit{UNETR}  \cite{UNETR} and \emph{SwinUNETR} \cite{SWIN_UNETR}, we split the dataset into 80 training and 20 testing samples following \cite{nnFormer}. The input is sampled at a resolution 96 x 96 x 96 with a voxel spacing of [1.5, 1.5, 2.0]. For \textit{nnFormer} \cite{nnFormer}, we follow their respective data processing pipeline and sample the input at a resolution 128 x 128 x 64 and report the results on all three classes using Dice similarity coefficient (DSC).
%
\textbf{BraTS Dataset:} We use two versions of BraTS dataset: BraTS17 \cite{brats17} and BraTS21 \cite{baid2021rsna}. 
% To be consistent with the baselines, we train UNETR \cite{UNETR} and SwinUNETR \cite{nnFormer} on BraTS21.
For \textit{UNETR, SwinUNTER} and \textit{PCRLv2} we report results on the BraTS21 dataset to be consistent with the baseline settings. 
The BraTS21 dataset \cite{baid2021rsna} is from the BraTS challenge and
% which provides a large dataset of 3D MRI scans, with voxel-wise ground truth labels annotated by clinicians. The dataset
includes 1251 subjects.
%each with four 3D MRI modalities: native T1, T1Gd, T2, and T2-FLAIR. 
% The dataset includes 1251 subjects, each with four 3D MRI modalities: native T1, post-contrast T1-weighted (T1Gd), T2-weighted (T2), and T2 Fluid-attenuated Inversion Recovery (T2-FLAIR).
% The images have been rigidly aligned, resampled to a 1 x 1 x 1 mm isotropic resolution, and skull-stripped, resulting in an input image size of 240 x 240 x 155.
% The dataset includes annotations for three tumor sub-regions: the enhancing tumor, the peritumoral edema, and the necrotic and non-enhancing tumor core. 
The annotations have been combined into three nested sub-regions: Whole Tumor (WT), Tumor Core (TC), and Enhancing Tumor (ET). Following the data split used by \cite{hatamizadeh2022swin}, we train on 1000 subjects and test on 251 subjects. 
% The input is cropped to a size 96 x 96 x 96 for training and Dice Similarity score (DSC) is used as an evaluation metric.
For \textit{nnFormer}, we use BraTS17 dataset \cite{brats17}. The task comprises of 484 MRI images.
%each having four channels - FLAIR, T1w, T1gd and T2w. 
% These images are obtained from 19 different institutions and represent a subset of the data used in the 2016 and 2017 Brain Tumor Segmentation (BraTS) challenges.
% The objective was to identify the three tumor sub-regions: edema (ED), enhancing tumor (ET), and non-enhancing tumor (NET). 
Following the dataset split of \cite{nnFormer}, we train on 387 training samples and test on 73 cases. 
% The input volumes are copped at a resolution of 128 x 128 x 64 for training. The evaluation metric used is (DSC).
Further details about the datasets and pre-processing are provided in Appendix \ref{sec:datasets-appendix}.

\textbf{Evaluation Metrics:}
% \label{subsec:eval-metrics}
To evaluate the models' performance, we utilize two metrics: the Dice Similarity Score (DSC) and the 95\% Hausdorff Distance (HD95). The DSC metric measures the degree of overlap between the volumetric segmentation predictions and the voxels of the ground truths as follows,
\begin{equation}
    {DSC}(Y,F) =  2 * \frac{ |Y \cap F|} {|Y| \cup |F|}=  2 * \frac{Y \cdot F} {Y^2 + F^2}
    \label{eq:DSC}
\end{equation}
where, $Y$ and $F$ denote the ground truths and output logits for all the voxels, respectively.

The HD95 metric is frequently employed as a boundary-based measure for determining the $95^{th}$ percentile of distances between the boundaries of the volumetric segmentation predictions and the voxels of the ground truths. Its definition is as follows:
\begin{equation}
    HD_{95}(Y,F) = \max \{ {d}_{YF}, {d}_{FY} \}
    \label{eq:HD95}
\end{equation}
Here, ${d}_{YF}$ represents the maximum distance at the $95^{th}$ percentile between the predicted voxels and the ground truth, while ${d}_{Y}$ represents the maximum distance at the $95^{th}$ percentile between the ground truth and the predicted voxels.

\textbf{Training and Implementation details:}
% \label{subsec:implementation}
Our approach utilizes Pytorch version 1.10.1 in conjunction with MONAI libraries \cite{monai} for implementation. To ensure fairness, we follow the respective training frameworks of the baseline architectures. Specifically, we use an input size of 128 x 128 x 64 for all datasets when training with \textit{nnFormer}, and 96 x 96 x 96 for \textit{UNETR}  and \textit{SwinUNTER}. All models are trained using a single A100 40GB GPU. For nnFormer, we train on all datasets for 1000 epochs, using AdamW optimizer \cite{loshchilov2017decoupled} with a learning rate of 0.01 and weight decay of 3e$^{-5}$, and for \textit{UNETR} and \textit{SwinUNETR}, we train for 5000 epochs on BTCV synapse, 1000 epochs on ACDC, and 300 epochs on BRaTs dataset, consistent with baseline settings.
The learning rate is kept default as per the given framework. 
During training, we apply the same data augmentations as used in UNETR, Swin UNETR, and nnFormer. Further details are available in the supplementary materials, and the code and models will be made publicly available.

% \subsection{Results}
% \label{subsec:results}
% \vspace{-0.2em}
\subsection{Comparison with state-of-the-art Baselines}
\label{subsubsec:  Comparison with State-of-the-art Baselines}
%As mentioned in Sec. \ref{sec:Introduction}, 
We show the effectiveness of our approach using three state-of-the-art 3D transformer based segmentation models: UNETR, SwinUNETR, nnFormer, and one CNN-based model: PCRLlv2,  across three datasets: Synapse Multi-Organ, ACDC, and BraTS (2017 and 2021).

\noindent
\textbf{Synapse Multi-Organ Dataset}:
Table \ref{table:synapse} shows the results on the synapse multi-organ dataset. We calculate performance metrics using Dice similarity score and HD95 score, maintaining consistent training settings as discussed above.
% We use Dice similarity score and HD95 score to calculate the performance metrics.The results are reported for the same number of epochs and training settings as mentioned in Sec. \ref{subsec:implementation}. 
UNETR with our approach achieves ~2.5\% higher Dice Score (81.13\%) than the baseline (78.76\%), and over 1\% reduction in HD95 score. With hierarchical SwinUNETR, Dice Score increases by $>$1\% (80.66\% to 82.00\%) and HD95 improves. For the complex nnFormer architecture, our approach enhances Dice score from 86.57\% to 87.35\%, with over 2\% HD95 reduction.
% With UNETR architecture integrated with our approach, we report a roughly 2.5\% increment in Dice Score (81.13\%) as compared to the baseline approach (78.76\%) and more than 1\% reduction in the error in terms of HD95 score. With SwinUNETR which is a hierarchical model, we observe  more than 1\% increment in the Dice Score (80.66\% to 82.00\%) and a decent error reduction in terms of HD95. For a more complex nnFormer architecture, we observe that our approach further complements the complex baseline architecture and improves the Dice score from 86.57\% to 87.35\%. Further, we observe a reduction of more than 2\% in the HD95 error. 
% \textit{Comparison with pretrained standard MIM \cite{xie2022simmim}: }We show the results on pretrained standard MIM approach in Table \ref{table:vanilla-mim-vs-ours}. Our 3D-MSR without any pretraining shows better results, demonstrating the superiority of our single stage training.
\textbf{ACDC Dataset}: Table \ref{table:ACDC} shows the results on the relatively bigger ACDC dataset
% that is a relatively bigger dataset than Synapse BTCV. 
% We use the Dice similarity score as the evaluation metric.
We observe that UNETR architecture complemented with our approach outperforms the baseline by a good margin of roughly 4\% in terms of Dice score (80.60\% vs 76.67\%). We further observe that the Dice  score per organ with our approach improves as compared to the baseline in each case. SwinUNETR shows over 2.5\% increment in Dice score over baseline, along with higher per organ dice scores. nnFormer follows a similar trend, achieving 90.73\% Dice score compared to the baseline's 90.50\%.
% With SwinUNETR, we observe more than 2.5\% increment in the Dice score. Also, the dice score per organ is higher with our approach. We observe a similar trend with nnFormer architecture, with a higher overall Dice score of 90.73\% as compared to the baseline with a dice score of 90.50\%. 
\textbf{BraTS Dataset}: 
%BraTS \cite{brats17, baid2021rsna} is a relatively larger dataset. 
 We use BraTS17 for UNETR and SwinUNETR architectures, and BraTS21 for nnFormer architecture. As shown in Table \ref{table:brats}, with UNETR architecture,  we report a gain of {0.54}\% in the overall dice score as compared to the baseline. With SwinUNTER, we report a Dice score of 89.83\% which is greater than baseline dice score of 89.44\%. Finally, with nnFormer, we observe a gain in the overall dice score (72.78\%) as compared to the baseline dice score (72.36\%). Overall we show that our approach achieves gains even on larger datasets such as BraTS, but has more pronounced improvement for the low-data setups.
% Further comparisons with other pretraining methods and masking schemes are provided in Appendix \ref{sec:pretrainingmim-vs-3dmsr-appendix} in the supplementary material.
% -- nnFormer -- BTCV Synapse (full, 5shot, 1shot), ACDC, MSD(Tumor)
% -- SwinUNETR -- BTCV Synapse (full, 5shot, 1shot), ACDC (running)
% -- UNETR -- BTCV Synapse (full, 5shot, 1shot), ACDC (running)

% \begin{table}[!t]
% \begin{minipage}{0.4\linewidth}
%  \centering \small\vspace{-1em}
%     \setlength{\tabcolsep}{3pt}
%     \scalebox{0.7}[0.7]{
%     \begin{tabular}{cc|a}
%     \toprule
%     \rowcolor{white}
%      Method & Pretraining & Average DSC\\
     
%     \midrule
%     \midrule
%     MIM %\cite{xie2022simmim}
%     & \cmark &79.67 \\
%     3D-MSR & \xmark &\textbf{81.13}  \\
%     \bottomrule
%     \end{tabular}}
%         \end{minipage}
%         \hfill
%         \begin{minipage}{0.5\linewidth}
%         \caption{\small Our 3D-MSR demonstrates superior performance as compared to pretrained MIM on Synapse with UNETR.}
%             % \caption{\small Our 3D-MSR demonstrates superior performance as compared to pretrained MIM on UNETR architecture across multiorgan synapse dataset.}
%    \label{table:vanilla-mim-vs-ours}
%         \end{minipage}
% \end{table}

\subsection{Few-shot performance}
\label{subsubsec: Few-shot performance}
We also validate the effectiveness of our approach in the few-shot scenario. Table \ref{table:few-shot} shows the performance comparison of our approach with the baselines in a 5-shot setting. We conduct few-shot experiments on Synapse BTCV  and ACDC datasets using all three model architectures.
% \cite{UNETR, SWIN_UNETR, nnFormer}. 
Specifically, on Synapse BTCV, our approach results in a 3-10\% increase in Dice score across all cases, indicating a significant improvement in segmentation accuracy. Similarly, on the relatively larger ACDC dataset, we observe a similar trend of higher Dice scores when our approach is integrated with the baseline models. This finding highlights the potential of our approach in improving the segmentation accuracy of medical images in situations where annotated data is limited, making it suitable for data-efficient training.

\begin{table}[!t]

\centering \small
\setlength{\tabcolsep}{4pt}
      \scalebox{0.58}[0.58]{
        \begin{tabular}{l l| c c c c c c c c c c c c c |a}
        \toprule
        \rowcolor{white}

            \multirow{2}{*}{\rotatebox[origin=c]{90}{Method}}  & \multirow{2}{*}{\rotatebox[origin=c]{90}{Pretrain}} & \multirow{2}{*}{\rotatebox[origin=c]{90}{Spleen}} &  \multirow{2}{*}{\rotatebox[origin=c]{90}{RKid}} &  \multirow{2}{*}{\rotatebox[origin=c]{90}{LKid}} & \multirow{2}{*}{\rotatebox[origin=c]{90}{Gall}}  & \multirow{2}{*}
            {\rotatebox[origin=c]{90}{Eso}}  & \multirow{2}{*}{\rotatebox[origin=c]{90}{Liv}}  & \multirow{2}{*}{\rotatebox[origin=c]{90}{Sto}} & \multirow{2}{*}
            {\rotatebox[origin=c]{90}{Aorta}} &  \multirow{2}{*}
            {\rotatebox[origin=c]{90}{IVC}} &  \multirow{2}{*}
            {\rotatebox[origin=c]{90}{Veins}} &  \multirow{2}{*}
            {\rotatebox[origin=c]{90}{Pan}} & \multirow{2}{*}
            {\rotatebox[origin=c]{90}{RAG}} & \multirow{2}{*}
            {\rotatebox[origin=c]{90}{LAG}}
            &  \multirow{2}{*}{\rotatebox[origin=c]{90}{Avg}}
            % \\ \cmidrule{16}
             % & & & & & & & & & & & & & & & \\
             \\
             \\
             \\
                \midrule
                % \midrule       
       Baseline & \xmark &89.0 & 89.2 & 87.7 & 47.6 & 48.9 & 94.4 & 74.7 & 82.0 &  77.3 & 61.7 &64.4 &56.6 &46.9 &70.8  \\
      \midrule
        SimCLR & \cmark &91.1  &91.3  &89.7 &48.7  &50.0  &96.6  &76.5 &83.9  &79.1   &  63.2 &65.9 &57.9 & 48.1&72.4
        \\
        % & \cmark & 87.22 &81.33  &83.55  &69.06 &95.31  &73.90  &90.10   &73.98  & \textbf{37.72} & \textbf{81.80} \\
        MAE & \cmark & 94.8 & 95.0 & 93.4 & 50.6&52.1 & 98.6  & 79.7 &  87.4&  82.4 & 65.9 & 68.6&60.5 &50.1 &75.3 \\
        SimMIM  & \cmark & 95.2 & 95.4 & 93.7 & 51.9 &52.3 & 98.7  & 79.9 &  87.7&  82.6 & 66.0 & 68.9&60.7 &51.2 &75.7 \\
        \midrule
        \textbf{MedContext} & \xmark & 93.8 & 93.7 & 93.6 & 54.9 &72.6 & 96.6  & 80.3 &  89.9&  83.3 & 72.9 & 73.9&64.4 &65.3 & \textbf{79.6} \\
        \bottomrule
        \end{tabular}}
        \caption{ MedContext vs. pretraining-finetuning \cite{Chen_2023_WACV}. DSC (\%) on Synapse dataset with UNETR architecture.}
         \label{table:pretrain-mim-vs-3dmsr}
         \vspace{-0.8em}
\end{table}

\begin{table}[!t]
\begin{minipage}{0.97\linewidth}
 \centering \small
    % \vspace{-1.5em}
 \setlength{\tabcolsep}{6pt}
        \scalebox{0.9}[0.9]{
        \begin{tabular}{cc|aaa}
        \toprule
        % \multirow{2}{*}{Method}  & \multicolumn{2}{c}{Average Dice Score} 
        % \\ \cmidrule{2-4}
         \rowcolor{white}
             Method  & Pretrain & Brats21 & ACDC & Synapse\\
        \midrule
        % \midrule
        PCRLv2 & \cmark &79.90 &78.53 &64.00\\
        PCRLv2 + MedContext & \xmark &\textbf{82.03}  &\textbf{82.57} &\textbf{72.30}\\
        \bottomrule
        \end{tabular}}
        % \end{minipage}
        % \hfill
        % \begin{minipage}{0.4\linewidth}
            % \caption{\scriptsize Comparison of 3D-MSR with PCRLv2 framework with 3D PCRLv2 architecture across 3 datasets.}
            % \caption{3D-MSR vs PCRLv2 framework with 3D PCRLv2 architecture across 3 datasets.}
            \caption{Improving PCRLv2 with our proposed MedContext across 3 datasets. We report Avg Dice scores.
            % using 3D PCRLv2 architecture.
            }
   \label{table:pcrlv2-vs-3dmsr}
        \end{minipage}
\vspace{-1.7em}
\end{table}

\subsection{Pretraining-Finetuning Baselines}
\label{subsubsec: Pretraining-Finetuning Baselines}
We demonstrate the effectiveness of MedContext by
comparing its performance with existing pretraining-finetuning methods in Table \ref{table:pretrain-mim-vs-3dmsr}.
% We observe that our 3D-MSR outperforms the methods which employ pretraining-finetuning paradigm without large-scale pretraining.
The baseline \cite{Chen_2023_WACV} method utilizes better weight initialization based on  pretraining with large dataset \cite{Harmon2020ArtificialIF} across state-of-the-art self-supervised methods \cite{chen2020simple, xie2022simmim, he2022masked} and then fine-tuned on the target dataset. In contrast, our MedContext directly learns the contextual cues from the target small dataset without the pre-training stage and outperforms the methods employing pretraining-finetuning paradigm. We further integrate our MedContext into the official implementation of PCRLv2 \cite{zhou2023pcrlv2}, a 3D CNN based state-of-the-art architecture, which is first pretrained in a self-supervised fashion on ~\cite{SETIO20171} and then finetuned on the specific target dataset. As seen in Table \ref{table:pcrlv2-vs-3dmsr}, our MedContext shows a complementary effect and improves the performance (Avg. Dice Score) of PCRLv2 architecture consistently without requiring pretraining. This also affirms the versatility of our framework, as it can be applied to CNN architectures, showcasing its universality.

% \begin{table}[!t]
% \begin{minipage}{0.4\linewidth}
%  \centering \small\vspace{-1em}
%     \setlength{\tabcolsep}{3pt}
%     \scalebox{0.7}[0.7]{
%     \begin{tabular}{cc|a}
%     \toprule
%     \rowcolor{white}
%      Method & Pretraining & Average DSC\\
     
%     \midrule
%     \midrule
%     MIM \cite{xie2022simmim} & \cmark &79.67 \\
%     3D-MSR & \xmark &\textbf{81.13}  \\
%     \bottomrule
%     \end{tabular}}
%         \end{minipage}
%         \hfill
%         \begin{minipage}{0.5\linewidth}
%         \caption{\small Our 3D-MSR demonstrates superior performance as compared to masked pretraining (MIM) with UNETR architecture on BTCV dataset.}
%             % \caption{\small Our 3D-MSR demonstrates superior performance as compared to pretrained MIM on UNETR architecture across multiorgan synapse dataset.}
%    \label{table:vanilla-mim-vs-ours}
%         \end{minipage}\vspace{-2em}
% \end{table}
\subsection{Ablation Studies}
\label{subsec:Ablation Studies}
\begin{figure}[!t]
\centering
  \begin{minipage}{.48\columnwidth}
  	\centering
    \includegraphics[ width=\linewidth,keepaspectratio]{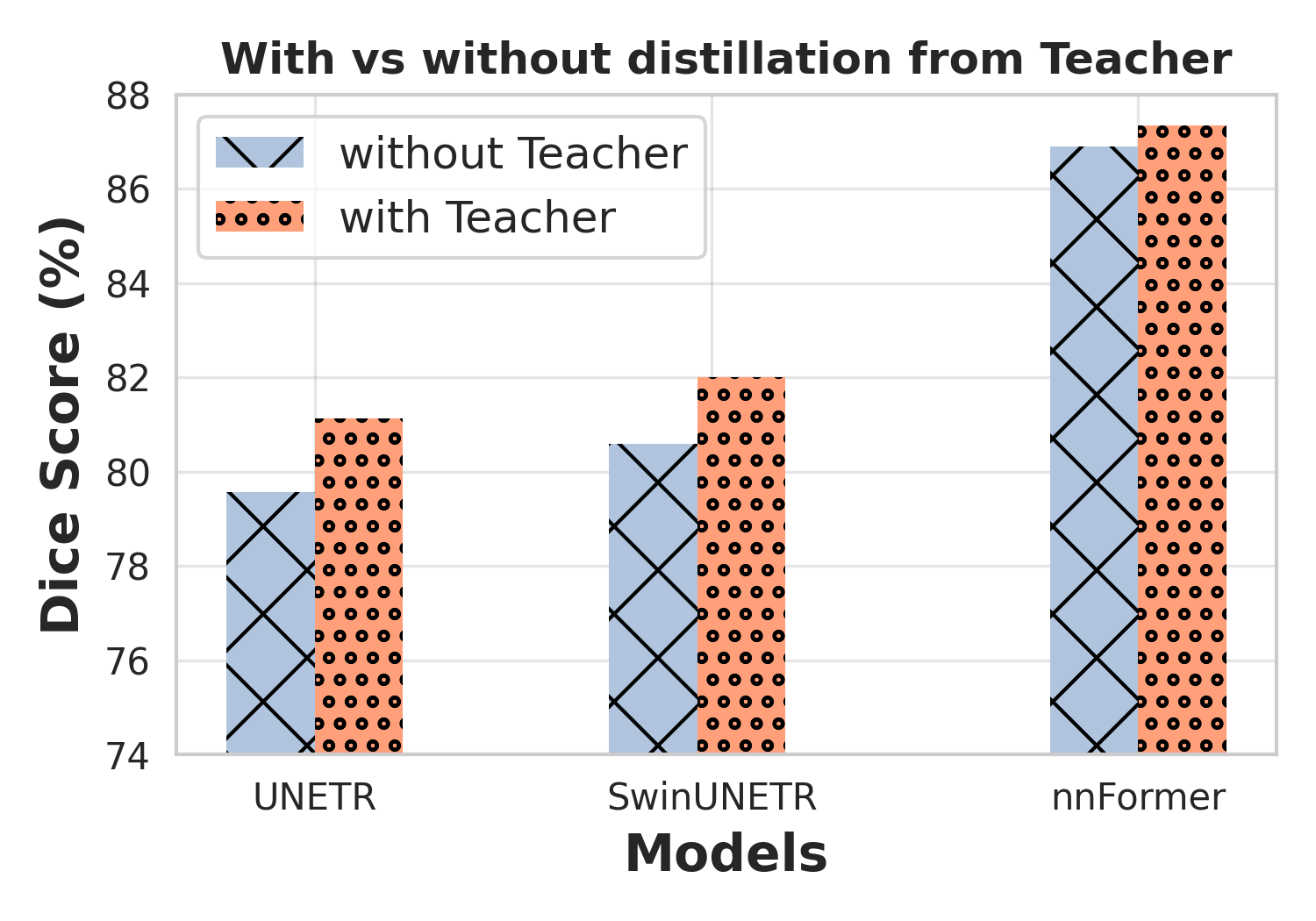}
  \end{minipage}
  \begin{minipage}{.48\columnwidth}
  	\centering
    \includegraphics[width=\linewidth, keepaspectratio]{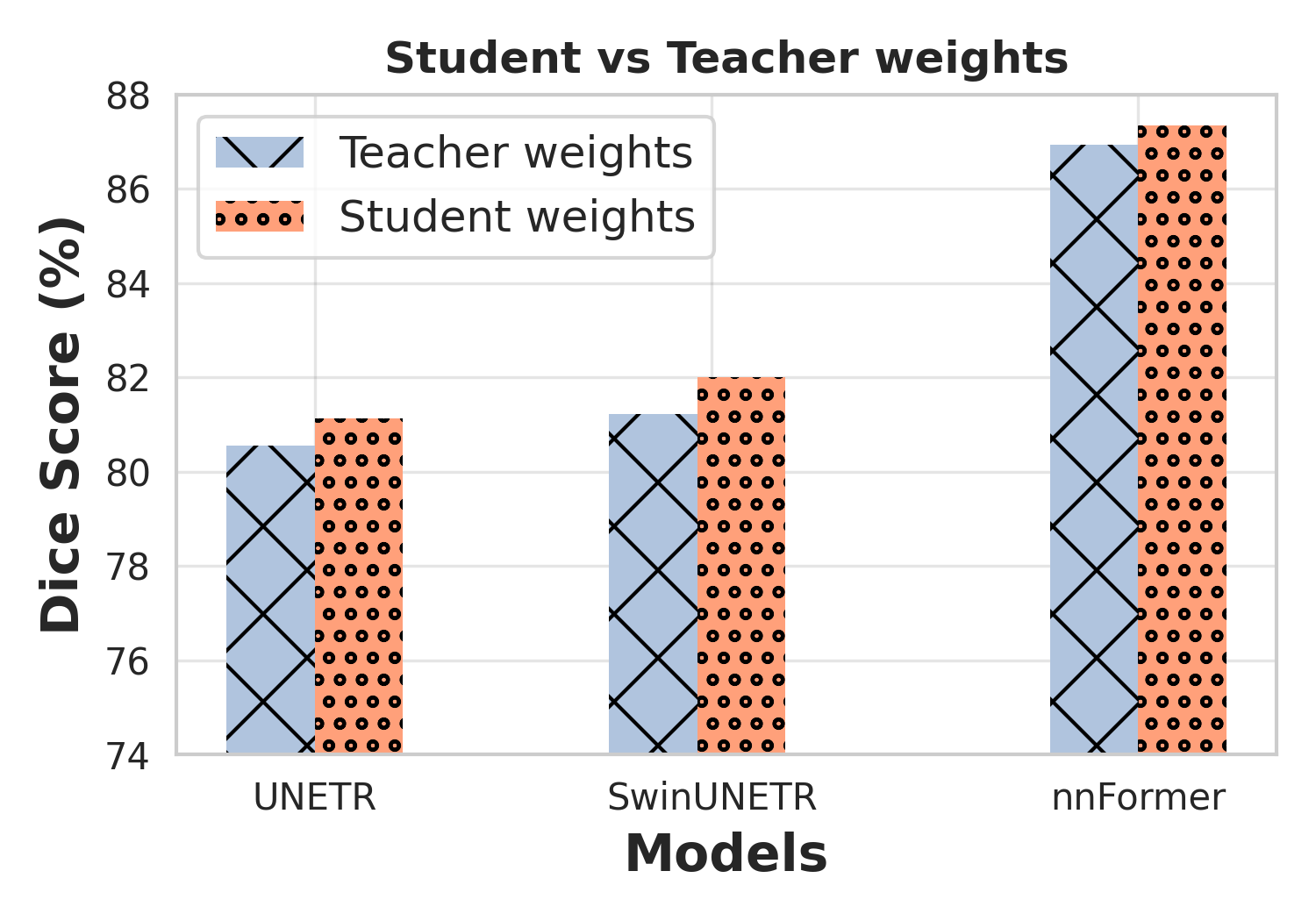}
  \end{minipage}%\vspace{-1em}
  \caption{DSC (\%) on Synapse across different models. \textbf{Left: Distillation from Teacher.} We demonstrate the importance of knowledge distillation through teacher for effectively leveraging contextual cues. \textbf{Right: Student vs Teacher.} We show that utilizing student weights during inference benefits overall performance. }
  \label{fig:with-vs-without-teacher}%\vspace{-1em}
  \vspace{-1em}
\end{figure}

\textbf{Effect of Student-Teacher framework: } As discussed in Sec. \ref{subsec:Voxelwise Segmentation Reconstruction from the Masked Input}, we hypothesize that the approach of training a single model on both original and masked input views using supervised loss with ground truth may not be sufficient for learning contextual relationships, as this method does not take into the account the knowledge acquired by the model during previous weights updates, which can lead to more effective guidance. This prompts us to use a student-teacher framework that leverages the information captured in the previous weight updates. We provide empirical evidence to support our claim in Fig. \ref{fig:with-vs-without-teacher} (left), where we report results on the Synapse multi-organ dataset across all three model architectures. In all cases, we observe a drop in performance in the absence of student-teacher framework.

\noindent
\textbf{Effect of Student vs Teacher weights:} Our proposed MedContext jointly optimizes the multi-task objectives in a  student-teacher framework. As discussed in Sec.~\ref{optimization}, the parameters of student and teacher networks are learned through different update rules. At inference, we can either choose the student or the teacher weights for predictions. We study the performance of student and teacher weights during inference in Fig.~\ref{fig:with-vs-without-teacher} (right). We observe that the student weights perform better than the teacher counterpart on synapse multi-organ dataset across two architectures.

% In each of the case, we observe that our approach integrated with the baseline outperforms the baseline by a decent margin. On Synapse BTCV, we observe that our approach integrated with the baselines gives approximately 3-6\% increment in Dice score in all the cases. Similarily, on a relatively larger ACDC dataset. we observe a similar trend where our approach gives higher Dice Score in all the cases. This establishes the fact that our approach is effective in the scenarios where data is scarce and hence can be used for data-efficient training.

\noindent
\textbf{Effect of Masking Ratio:} We propose to learn the contextual knowledge by allowing the model to reconstruct the missing regions of the masked input in the segmentation space. However, the number of missing patches to recover may influence the performance of the model. We conduct an ablation study on a held-out validation set to determine the optimal masking ratio. %for the objective of contextual consistency. 
Table \ref{table:ablations-masking} shows the Dice scores for a range of masking ratios across Synapse dataset using UNETR and SwinUNETR architectures complemented with our approach. 
% Our approach consistently betters results, with 40\% masking ratio showing optimal improvement.
Although our method shows improvement on all masking ratios, however, we observe that a 40\% masking ratio works best in our case. 

\noindent
\textbf{Effect of different losses:} Our proposed approach utilizes multiple supervised and self-supervised losses in the training stage. As discussed in Secs. \ref{subsec:Voxelwise Segmentation Reconstruction from the Masked Input} and \ref{supervised-segmentation}, each loss component holds a specific significance towards the final objective.
% We study the importance of each loss component in Table \ref{table:ablations-loss-components}.
We conduct an ablative analysis on BTCV Synapse dataset to study the importance of the 2 loss components: Masked student loss (MSL) and  Consistency Loss (CL), across two architectures: UNETR and SwinUNETR, in Table \ref{table:ablations-loss-components}. We observe that removing any one of the loss components results in a drop in the Dice score. 
% We notice the trend in both architectures, offering empirical evidence that the mutual synergy between supervised and self-supervised losses enhances contextual cues for effective 3D medical segmentation.
We observe this trend with both the architectures, providing empirical evidence
that mutual synergy between supervised and self-supervised losses helps induce contextual cues for effective 3D medical segmentation.

\begin{table}[!t]
\begin{minipage}{0.4\linewidth}
 \centering \small 
 \setlength{\tabcolsep}{5pt}
        \scalebox{0.7}[0.6]{
        \begin{tabular}{c|aa}
        \toprule
        \multirow{2}{*}{Masking ratio}  & \multicolumn{2}{c}{Average Dice Score} 
        \\ \cmidrule{2-3}
         \rowcolor{white}
               &UNETR &SwinUNETR\\
        \midrule
        \midrule
        30\% &79.54  &80.92\\
        40\% & \textbf{80.47} &\textbf{82.00}\\
        50\% & 80.00  &81.03\\ 
        60\% & 80.20 &81.70\\
        80\% & 79.90  &81.27\\
        \bottomrule
        \end{tabular}}
        \end{minipage}
        \hfill
        \begin{minipage}{0.48\linewidth}
            \caption{\small Understanding the effect of masking ratio on our MedContext. We report the Dice score on the Synapse across UNETR \& SwinUNETR architectures.
            % The best results are shown in bold.
            }
            % \caption{\small We demonstrate the impact of the masking ratio on the segmentation performance. We report the Dice score on Synapse dataset across UNETR and SwinUNETR architectures. 
            % % The best results are shown in bold.
            % }
   \label{table:ablations-masking}
        \end{minipage} 
        \vspace{-1.5em}
\end{table}

\begin{table}[!t]
\begin{minipage}{0.35\linewidth}
 \centering \small %\vspace{-1em}
    \setlength{\tabcolsep}{3pt}
    \scalebox{0.85}[0.75]{
    \begin{tabular}{cc|aa}
    \toprule
     MSL & CL & \multicolumn{2}{c}{Average Dice Score}\\
      \cmidrule{3-4}
     \rowcolor{white}
    & &  UNETR &SwinUNETR\\

    \midrule
    \midrule
    
     \cmark  &\xmark  &78.69 &81.03 \\
    \xmark  &\cmark  &79.46  &81.25\\
    \cmark  &\cmark  &\textbf{80.32}  & \textbf{81.70}\\
    \bottomrule
    \end{tabular}}
        \end{minipage}
        \hfill
        \begin{minipage}{0.48\linewidth}
            \caption{\small We show the effect of each loss component on final objective (Eq.~\ref{eqn:combined-loss}). We report dice score (\%) on Synapse  dataset.}% for UNETR and SwinUNETR .}
   \label{table:ablations-loss-components}
        \end{minipage}
\vspace{-1.5em}
\end{table}

\section{Conclusion}
In this paper, we propose a universal training framework called \textit{MedContext} which effectively learns self-supervised contextual cues jointly with the supervised voxel segmentation task without requiring large-scale annotated volumetric medical data. Our proposed approach employs a student-teacher distillation strategy to reconstruct missing organs or parts of organs in the output segmentation space. Through extensive experimentation, our approach demonstrates complementary benefits to existing state-of-the-art 3D medical segmentation architectures in both conventional and few-shot settings without pretraining on large-scale datasets. Moreover, the plug-and-play design of our approach allows for its easy integration into any architectural design.
{
    \small
    \bibliographystyle{ieeenat_fullname}
    \bibliography{references}
}

\newpage
\clearpage
 % \clearpage
% \setcounter{page}{1}
% \maketitlesupplementary

\appendix
\noindent\begin{huge} \textbf{Supplementary Material} \vspace{4mm} \end{huge}

In this section we discuss the Psuedocode of our MedContext Algorithm in Appendix \ref{sec:algorithm-appendix}, provide additional details about the datasets in Appendix \ref{sec:datasets-appendix}, 
% show further quantitative comparisons with existing work of \citesupp{suppChen_2023_WACV} in Appendix \ref{sec:pretrainingmim-vs-3dmsr-appendix} and
% n medical segmentation utilising large pretraining dataset. 
finally, in Appendix \ref{sec:qualitative-examples}, we provide some qualitative visualizations to show the superiority of our MedContext for effective 3D medical segmentation.

\section{MedContext Algorithm}
\label{sec:algorithm-appendix}
In section \ref{sec: Architecture} of the main paper, we describe our MedContext algorithm which comprises a student network ($F_s$) and a teacher network ($F_t$). The student-teacher framework processes the input volume $\bm{X} \in \mathbb{R}^{H\times W \times D}$ and its masked version $\bm{X}^M \in \mathbb{R}^{H\times W \times D}$ created using the masking function $g(.)$. During training, the student-teacher framework generates voxel-wise semantic logits for each input view. The student network receives both the masked ($\bm{X}^M$) and unmasked ($\bm{X}$) inputs, and outputs voxel-wise semantic logits represented by $\bm{F}_{s}$ and $\bm{F}_{s}^{M}$. The teacher network receives only the original unmasked input $\bm{X}$ and generates voxel-wise semantic logits denoted by $\bm{F}_{t}$. We calculate a supervised loss using the ground truth label $\bm{Y}$ for each output prediction from student. Additionally, we minimize a self-supervised objective between the masked student logits $\bm{F}_{s}^{M}$ and the teacher logits $\bm{F}_{t}$. The hyperparameter $\beta$ controls the contribution of self-supervised loss to the overall objective function. Our single-stage training process jointly optimizes both the supervised and self-supervised objectives. At the inference stage, we use the weights of the student network to output the corresponding voxel-wise predictions. We provide the Pseudo-code of our MedContext approach in Algorithms \ref{alg:training} and \ref{alg:inference}.

\begin{algorithm}[h]\small
   \caption{MedContext: Training}
   \label{alg:training}
\begin{algorithmic}
   \STATE {\bfseries Input:} Dataset $D$, student $\mathcal{F}_{\theta_{s}}$, teacher $\mathcal{F}_{\theta_{t}}$, masking function $g(.)$ with masking ratio $\alpha$, Step = 0. 
   \STATE {\bfseries Require:} Initialize teacher weights $\theta_{t}$ with the student weights $\theta_{s}$. Network momentum rate $m$ follows a cosine schedule from 0.996 to 1. Soft Dice loss $\mathcal{L}_{Dice-CE}$. \\
   \REPEAT
   \STATE Step $\leftarrow$ Step + 1 \\
   \tcp{Sample data.}
   \STATE sample  $\{X, Y\} \subseteq D$, $X^M \leftarrow g(X,\alpha)$\\
   \tcp{student and teacher outputs.}
   \STATE $F_s$ , $F_s^M$ =  $\mathcal{F}_{\theta_{s}}(X)$ , $\mathcal{F}_{\theta_{s}}(X^M)$   \\
   \STATE $F_t$  =  $\mathcal{F}_{\theta_{t}}(X)$   \\
   \tcp{supervised objective}
   \STATE $\mathcal{L}(Y, F_{s}) \leftarrow  \mathcal{L}_{Dice-CE}(Y, F_s)$   \\ 
   % \STATE $\mathcal{L}(Y, F_{t}) \leftarrow  \mathcal{L}_{Dice-CE}(Y, F_t)$   \\
   \tcp{masked reconstruction objectives.}
    \STATE $\mathcal{L}(Y, F_{s}^M) \leftarrow  \mathcal{L}_{Dice-CE}(Y, F_s^M)$   \\ 
   \STATE $\mathcal{L}_{c}(F_t, F_{s}^M) \leftarrow \frac{\Vert\ F_s^M - F_t  \Vert\ _{2}^{2}}{\Vert\ F_t  \Vert\ _{2}^{2}} $   \\
   \tcp{Combined Loss.}
   \STATE $\mathcal{L} \leftarrow \mathcal{L}(Y, F_{s}) + \mathcal{L}(Y, F_{s}^M) + \beta \mathcal{L}_{c}(F_t, F_{s}^M) $ \\
   \tcp{update student with combined loss.}
    $\theta_s \leftarrow  \theta_s - \delta \nabla_{\theta_s} (\mathcal{L})$ \\
   \tcp{update teacher weights by EMA.}
    $\theta_t \leftarrow  m \theta_t - (1-m)\theta_s$ \\
    \UNTIL{converge}
\end{algorithmic}
\end{algorithm}

\begin{algorithm}[ht!]\small
   \caption{Inference}
   \label{alg:inference}
\begin{algorithmic}
   \STATE {\bfseries Input:} Test Dataset $D'$, student network $\mathcal{F}_{\theta_{s}}$ initialized with learned paramaters $\theta_{s}$, Step = 0. 
   \STATE {\bfseries Require:} Evaluation metric $\mathcal{E}$. \\
   \REPEAT
   \STATE Step $\leftarrow$ Step + 1 \\
   \tcp{Sample data.}
   \STATE sample  $\{X, Y\} \subseteq D'$\\
   \tcp{model prediction.}
   \STATE $F$ =  $\mathcal{F}_{\theta_{s}}(X)$   \\
   \tcp{calculate evaluation metric.}
   \STATE Evaluate $ \leftarrow  \mathcal{E}(Y, F)$   \\ 
    \UNTIL{go through all test data}
\end{algorithmic}

\end{algorithm}

\section{Additional Dataset details}
\label{sec:datasets-appendix}
We use three medical segmnetation datasets in our paper. The additional details of each dataset along with the preprocessing settings are given below:\\
\textbf{Synapse BTCV Multi-organ Dataset \cite{BTCV}:}  
The BTCV dataset, known as Synapse for Multi-organ CT Segmentation, is derived from the MICCAI Multi-Atlas Labeling Beyond the Cranial Vault challenge. It includes abdominal CT scans of 30 subjects encompassing 8 organs. The dataset is expertly annotated under the supervision of clinical radiologists at Vanderbilt University Medical Center. Each scan is captured using contrast enhancement in the portal venous phase and contains between 80 to 225 slices with 512×512 pixels. The thickness of each slice varies from 1 to 6 mm. 
Following previous methods, we adopt the same dataset split as used in \cite{TransUNet} with 18 train samples and test on the remaining 12 cases. We evaluate the performance on eight abdominal organs (i.e., spleen, right kidney, left kidney, gallbladder, liver, stomach, aorta, and pancreas) by measuring the Dice Similarity Coefficient (DSC) and 95\% Hausdorff Distance (HD95). In all the cases, the intensities of input volumes are normalized from the range of [-1000, 1000] to [0,1] Hounsfield Units (HU). 
For \textit{UNETR}  \cite{UNETR} and \emph{SwinUNETR} \cite{SWIN_UNETR}, we follow their respective data processing pipeline where each volume is pre-processed independently and resampled to have an isotropic voxel spacing of [1.5, 1.5, 2.0]. Input is sampled at a crop size of 96 x 96 x 96. For \textit{nnFormer} \cite{nnFormer}, we use their data preprocessing pipeline in which each CT scan is independently processed by applying patch cropping to sample the input at a resolution 128 × 128 × 64 with a spacing of [0.76, 0.76, 3].

\textbf{ACDC Dataset \cite{ACDC}:} The ACDC dataset is a collection of cardiac MRI images and associated segmentation annotations for the right ventricle (RV), left ventricle (LV), and myocardium (MYO) of 100 patients, obtained from actual clinical exams. The dataset includes patients with various heart conditions, including normal patients, patients with a myocardial infarction, dilated cardiomyopathy, hypertrophic cardiomyopathy, and abnormal right ventricle. 
For  \textit{UNETR}  
%\cite{UNETR} 
and \emph{SwinUNETR},\cite{SWIN_UNETR},
we split the dataset into 80 training and 20 testing samples following \cite{nnFormer}. The input is sampled at a resolution 96 x 96 x 96 with a voxel spacing of [1.5, 1.5, 2.0]. For \textit{nnFormer \cite{nnFormer}}, we follow their respective data processing pipeline and sample the input at a resolution 128 x 128 x 64 and report the results on all three classes using Dice similarity coefficient (DSC).

\textbf{BraTS Dataset \cite{brats17,baid2021rsna}}: We use two versions of BraTS dataset: BraTS17 \cite{brats17} and BraTS21 \cite{baid2021rsna}. 
% To be consistent with the baselines, we train UNETR \cite{UNETR} and SwinUNETR \cite{nnFormer} on BraTS21.
For \textit{UNETR and SwinUNTER} we report results on the BraTS21 dataset to be consistent with the baseline settings. The BraTS21 dataset \cite{baid2021rsna} is from the BraTS challenge which provides a large dataset of 3D MRI scans, with voxel-wise ground truth labels annotated by clinicians.
The dataset includes 1251 subjects, each with four 3D MRI modalities: native T1, post-contrast T1-weighted (T1Gd), T2-weighted (T2), and T2 Fluid-attenuated Inversion Recovery (T2-FLAIR).
The images have been rigidly aligned, resampled to a 1 x 1 x 1 mm isotropic resolution, and skull-stripped, resulting in an input image size of 240 x 240 x 155.
The dataset includes annotations for three tumor sub-regions: the enhancing tumor, the peritumoral edema, and the necrotic and non-enhancing tumor core. 
The annotations have been combined into three nested sub-regions: Whole Tumor (WT), Tumor Core (TC), and Enhancing Tumor (ET). 
Following the data split used by \cite{SWIN_UNETR}, we train on 1000 subjects and test on 251 subjects. The input is cropped to a size 96 x 96 x 96 for training and Dice Similarity score (DSC) is used as an evaluation metric.
For \textit{nnFormer}, we use BraTS17 dataset \cite{brats17}. The task comprises of 484 MRI images, each having four channels - FLAIR, T1w, T1gd and T2w. 
These images are obtained from 19 different institutions and represent a subset of the data used in the 2016 and 2017 Brain Tumor Segmentation (BraTS) challenges.
The objective was to identify the three tumor sub-regions: edema (ED), enhancing tumor (ET), and non-enhancing tumor (NET).

% \section{Comparison with pretraining approaches}
% \label{sec:pretrainingmim-vs-3dmsr-appendix}
% We demonstrate the effectiveness of our 3D-MSR by comparing its performance with existing work by Chen  \citesupp{suppChen_2023_WACV}, which utilizes a large dataset of 771 CT volumes for pretraining using different self-supervised approaches \citesupp{suppchen2020simple, supphe2022masked, suppxie2022simmim}. We compare the performance in terms of Dice similarity score (\%) using the UNETR architecture across the synapse multi-organ dataset on all the 13 organs. We follow the same dataset split of Chen et al. \citesupp{suppChen_2023_WACV} and take the first 24 volumes for training and the remaining 6 for validation. We report the scores on the validation set. We observe that the supervised baseline without any pretraining performs the worst among all the methods, which is expected. Constrastive pretraining on large dataset with SimCLR boosts the performance of the baseline, however, the gain is not as much pronounced as compared to masking-based schemes such as MAE \citesupp{supphe2022masked} and SimMIM \citesupp{suppxie2022simmim}.  Compared to the supervised baseline, our 3D-MSR produces a gain of approximately 9\% in Dice similarity score. Similarly, compared to the contrastive and masking based pretraining methods, our 3D-MSR without any pretraining outperforms all the approaches as shown in Table \ref{table:appendix-mim-vs-3dmsr} by a margin of approximately 4-7\%.

\section{Qualitative Comparisons}
\label{sec:qualitative-examples}
% We showcase the qualitative comparison of our 3D-MSR with baseline on synapase multi-organ dataset across UNETR architecture in Figure \ref{fig:qualitative_synapse_unetr}. As seen from the figure, our 3D-MSR produces accurate segmentations as compared to baseline aproach. In the second row of Fig. \ref{fig:qualitative_synapse_unetr}, the baseline UNETR struggles to segment the pancreas as marked in red rectangle. After integrating the baseline with 3D-MSR, the model correctly segments the pancreas organ (second row, fourth column). Similarly in the third row, the baseline model segments half of the right adrenal gland (shown in yellow) as spleen, which is correctly segmented in our approach.

\begin{figure*}[!t]
% \vspace{-1.0em}
\centering
  \begin{minipage}{0.8\linewidth}
  	\centering
    \includegraphics[width=0.98\linewidth,keepaspectratio]{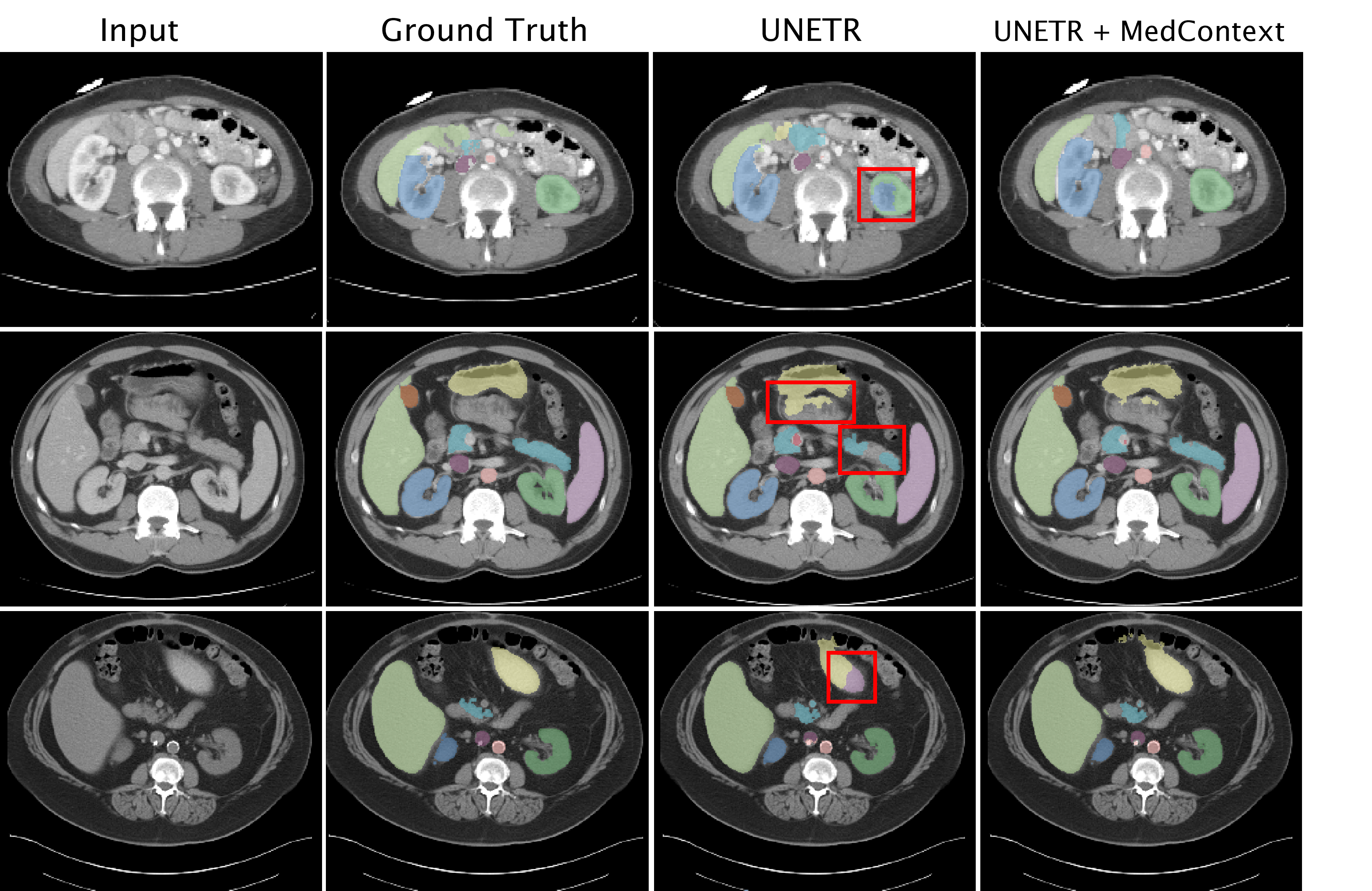}
   \end{minipage}
  \begin{minipage}{0.85\linewidth}
  
\scalebox{0.06}{{\usebox{\spleen}}} \small Spleen ~~\scalebox{0.06}{{\usebox{\rkid}}} \small R-Kid ~~\scalebox{0.06}{{\usebox{\lkid}}} \small L-Kid ~~\scalebox{0.06}{{\usebox{\gall}}} \small Gal~~\scalebox{0.06}{{\usebox{\eso}}} \small Eso ~~\scalebox{0.06}{{\usebox{\liver}}} \small Liv ~ ~\scalebox{0.06}{{\usebox{\sto}}} \small Sto ~~\scalebox{0.06}{{\usebox{\aorta}}} \small Aor ~~\scalebox{0.06}{{\usebox{\ivc}}} \small ICV ~~\scalebox{0.06}{{\usebox{\veins}}} \small PSV ~~\scalebox{0.06}{{\usebox{\panc}}} \small Pan ~~\scalebox{0.06}{{\usebox{\rad}}} \small Rad. ~~\scalebox{0.06}{{\usebox{\lad}}} \small Lad.
\end{minipage}
% \vspace{-0.5em}
% \hfill
%   \begin{minipage}{0.48\linewidth}
%   	\centering
%     \includegraphics[ width=\linewidth,keepaspectratio]{3D-MedicalSegmentation-ICCV2023/figures/nnformer_acdc_vis.png}
%   % \end{minipage}
%   % \begin{minipage}{\linewidth}
% \scalebox{0.06}{{\usebox{\rad}}} \small RV cavity ~~~~~~~~~ \scalebox{0.06}{{\usebox{\lkid}}} \small Myocaridum ~~~~~~~~~~~~
% \scalebox{0.06}{{\usebox{\RV}}} \small LV cavity 
% \end{minipage}
  \caption{Qualitative comparison on multi-organ synapse dataset: We  showcase the benefit of our MedContext framework implemented on the UNETR architecture. The examples display various abdominal organs, with their corresponding labels in the legend below. The existing baseline method struggles to accurately segment the organs as can be seen from the red boxes. Best viewed in zoom. } 
  \label{fig:qualitative_synapse_unetr}
    % \vspace{0.4em}
\end{figure*}

% Figure \ref{fig:qualitative_acdc_unetr} shows the qualitative comparison of our 3D-MSR approach integrated with UNETR \cite{UNETR} on ACDC dataset \cite{ACDC}. We observe that our 3D-MSR integrated with UNETR produces sharp and correct segmentation boundaries as compared to baseline UNETR. As for instance, in the third row of Figure \ref{fig:qualitative_acdc_unetr}, the baseline completely misses the right ventricular (RV) cavity which is segmented efficiently with our approach.
In Figure \ref{fig:qualitative_synapse_unetr}, we present a comparison between our MedContext approach and the baseline method on the synapse multi-organ dataset using the UNETR architecture. The results indicate that our approach produces more accurate segmentations than the baseline. As for instance, in the first row of Figure \ref{fig:qualitative_synapse_unetr}, the baseline (first row, third column) wrongly segments the right kidney (dark green) as left kidney (dark blue). Our approach on the other hand segments both the kidneys correctly (first row, fourth column). In the second row of the figure, for example, the baseline struggles to segment the pancreas, but when combined with our MedContext approach, the model correctly segments the organ. Similarly, in the third row, our approach correctly segments the right adrenal gland, which the baseline model almost wrongly segments as spleen.

\begin{figure*}[!t]
 % \vspace{-1.0em}
\centering
  \begin{minipage}{0.7\linewidth}
  	\centering
    \includegraphics[ width=\linewidth,keepaspectratio]{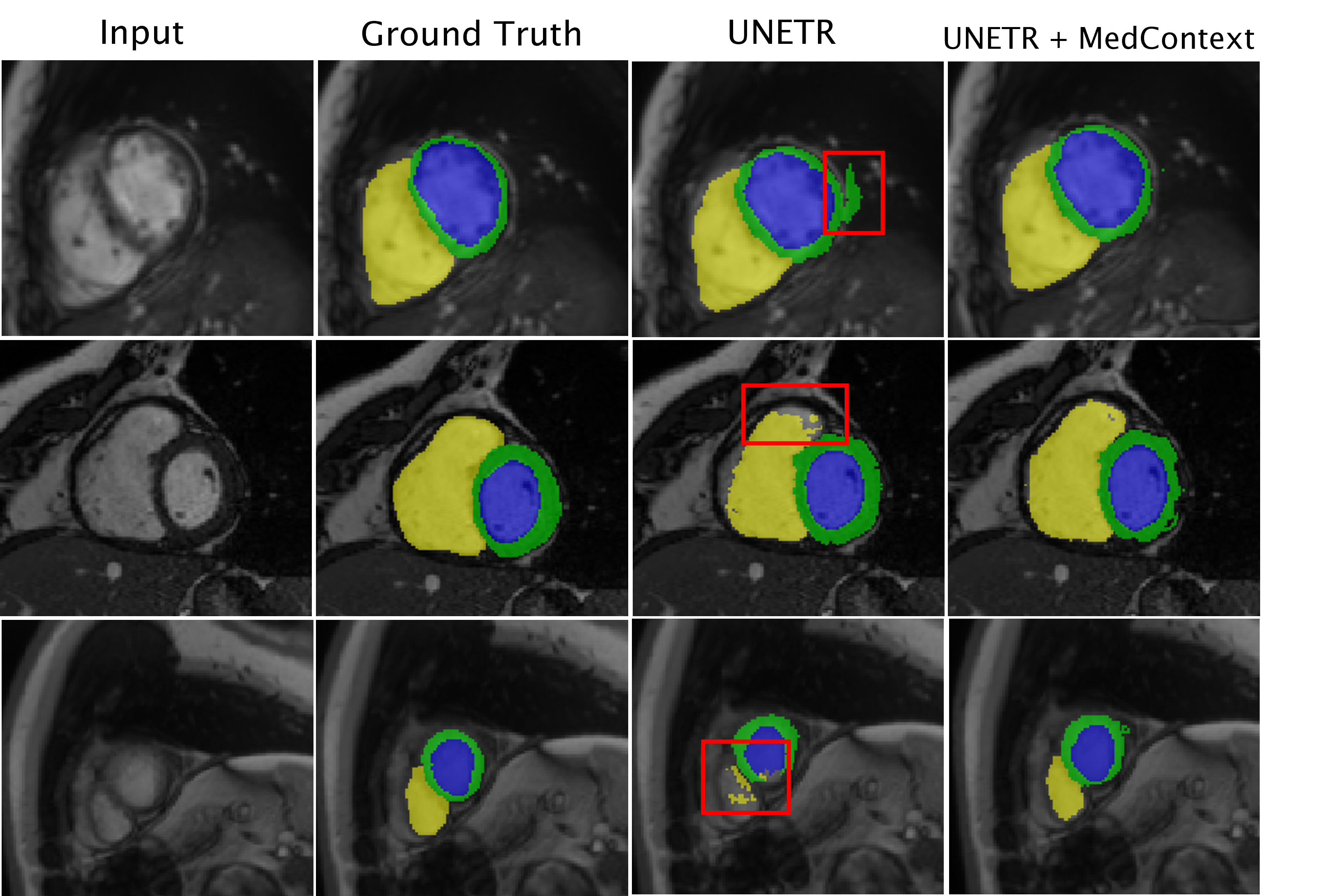}
  % \end{minipage}
  % \begin{minipage}{\linewidth}
\scalebox{0.06}{{\usebox{\rad}}} \small RV cavity ~~~~~~~~~ \scalebox{0.06}{{\usebox{\lkid}}} \small Myocaridum ~~~~~~~~~~~~
\scalebox{0.06}{{\usebox{\RV}}} \small LV cavity
\end{minipage}
% \vspace{-0.8em}
% \hfill
%   \begin{minipage}{0.48\linewidth}
%   	\centering
%     \includegraphics[ width=\linewidth,keepaspectratio]{3D-MedicalSegmentation-ICCV2023/figures/nnformer_acdc_vis.png}
%   % \end{minipage}
%   % \begin{minipage}{\linewidth}
% \scalebox{0.06}{{\usebox{\rad}}} \small RV cavity ~~~~~~~~~ \scalebox{0.06}{{\usebox{\lkid}}} \small Myocaridum ~~~~~~~~~~~~
% \scalebox{0.06}{{\usebox{\RV}}} \small LV cavity 
% \end{minipage}
  \caption{Qualitative comparison on ACDC dataset using UNETR: We  showcase the benefit of our MedContext framework integrated with UNETR architecture on ACDC dataset. The examples display three heart regions with their corresponding labels in the legend below. The baseline UNETR struggles to accurately segment the organs as can be seen from the red boxes. Our approach on the other hand produces correct and sharp segmentation boundaries. Best viewed in zoom. }
  \label{fig:qualitative_acdc_unetr}

\end{figure*}

% Similarly, in Figure \ref{fig:qualitative_acdc_unetr}, we show the results of our 3D-MSR approach integrated with UNETR on the ACDC dataset. The comparison indicates that our approach produces sharper and more accurate segmentation boundaries than the baseline UNETR method. For instance, the segmentation boundary of Myocardium (shown in green)  for the baseline (first row, third column)  deviates from the given ground truth segmentation boundary. We see the similar trend in the second row, third column where the baseline fails to efficiently encompass the segmentation boundary of Myocardium. Similarly, in the third row, baseline completely misses the right ventricular cavity which our approach segments correctly.
Similarly, Figure \ref{fig:qualitative_acdc_unetr} illustrates the outcomes of integrating our MedContext technique with UNETR on the ACDC dataset. Our method produces segmentation boundaries that are more precise and well-defined than the baseline UNETR method. The baseline's segmentation boundary of the Myocardium (shown in green) in the first row's third column is not accurate when compared to the ground truth segmentation boundary. This trend is also observed in the second row's third column where the baseline fails to segment the Myocardium efficiently. Additionally, in the third row, the baseline fails to identify the right ventricular cavity. In contrast, our MedContext outperforms the baseline in all the cases and efficiently segments all the organs.

% {\small
% \bibliographystyle{ieee_fullname}
% \newpage
% { \small
%     \bibliographystylesupp{ieeenat_fullname}
%     \bibliographysupp{suppbib}
% }

\end{document}